\documentclass[iop, apj]{emulateapj}

\newcommand{\bo}{\mbox{\boldmath $\omega$}}
\newcommand{\bmu}{\mbox{\boldmath $\mu$}}
\newcommand{\bl}{\mbox{\boldmath $\ell$}}
\newcommand{\blhat}{\mbox{\boldmath $\hat \ell$}}
\newcommand{\bohat}{\mbox{\boldmath $\hat \omega$}}
\newcommand{\bxhat}{\mbox{\boldmath ${\hat x}$}}
\newcommand{\byhat}{\mbox{\boldmath ${\hat y}$}}
\newcommand{\bzhat}{\mbox{\boldmath ${\hat z}$}}
\newcommand{\Porb}{P_{\rm orb}}

\shorttitle{The Emergence of Negative Superhumps in CVs}
\shortauthors{Thomas \& Wood}

\begin{document}

\title{The Emergence of Negative Superhumps in Cataclysmic Variables:\\ Smoothed Particle Hydrodynamics Simulations}

\author{David~M.~Thomas}
\affil{Department of Physics and Space Sciences, Florida Institute of Technology, \\
    150 W University Blvd., Melbourne FL 32901 USA}
    \and
\author{Matt~A.~Wood}
\affil{Department of Physics and Astronomy, Texas A\&M University--Commerce,\\
    Commerce, TX 75429-3011, USA}
    
\email{dthomas@my.fit.edu}
\email{Matt.Wood@tamuc.edu}	

\begin{abstract}
Negative superhumps are believed to arise in cataclysmic variable systems when the accretion disk is tilted with respect to the orbital plane. Slow retrograde precession of the line-of-nodes results in a signal---the negative superhump---with a period slightly less than the orbital period. Previous studies have shown that tilted disks exhibit negative superhumps, but a consensus on how a disk initially tilts has not been reached. Analytical work by Lai suggests that a magnetic field on the primary can lead to a tilt instability in a disk when the dipole moment is offset in angle from the spin axis of the primary and when the primary's spin axis is, itself, not aligned with the angular momentum axis of the binary orbit. However, Lai did not apply his work to the formation of negative superhumps.  In this paper, we add Lai's model to an existing smoothed particle hydrodynamics code. Using this code, we demonstrate the emergence of negative superhumps in the ``light curve'' for a range of magnetic dipole moments.  We show that the period deficits calculated from these negative superhumps match those in simulations using manually tilted disks. When positive superhumps appear ($q \lesssim 0.33$), we show that the period excesses calculated from these signals are also consistent with previous results. Using examples, we show that the disks are tilted, though the tilt varies periodically, and that they precess in the retrograde direction. The magnetic fields found to lead to the emergence of negative superhumps lie in the kilogauss regime.  

\end{abstract}


\section{Introduction}

Cataclysmic variables (CVs) are binary star systems containing a white dwarf (WD) primary and (typically) a main-sequence companion star. The secondary, though less massive than the WD, fills its Roche lobe.  As a result, material flows from the secondary through the inner Lagrange point $L_1$ and onto the primary.  Except for the highly-magnetic CVs and the two AM CVn direct impact systems, conservation of momentum requires that a disk form about the primary.  The disk provides a mechanism (via viscosity) for transporting mass onto the primary (accretion) while allowing angular momentum to move outward, and is the principal source of the variations in the system's luminosity.  CVs are broadly divided into two major classes, the non-magnetic systems and the magnetic systems. 
The classification ``non-magnetic'' is generally taken to indicate that Zeeman-split spectral lines are not observed, but this observation becomes difficulty below $\sim$$10^5$ G, corresponding to a magnetic dipole moment of $\sim$$10^{31}\ \rm G\ cm^3$. It is widely assumed that many ``non-magnetic" systems do in fact have magnetic fields strong enough ($\sim$$10^3$---$10^5$G)  to affect the plasma flow near the accreting white dwarf without making the star observably magnetic \citep[e.g.,][]{1998PASP..110..403P,warner2004}, a distinction that is important for the work presented here.  The non-magnetic systems traditionally include the SU UMa, U Gem, and Z Cam subclasses.  Polars (also known as AM Her stars) and intermediate polars (alternately, DQ Her stars) comprise the magnetic stars.  Polars have magnetic fields strong enough ($\gtrsim 10$ MG) that no accretion disk is formed -- the accretion stream originating in the $L_1$ region flows ballistically to a broad transition region, and then follows the field lines to the magnetic poles of the white dwarf.  Intermediate polars have weaker fields ($\sim$1--$10$ MG), allowing an outer disk to form, the inner boundary of which is a transition region located at a radius determined by the field strength and accretion rate, and then accretion via magnetic curtains along the field lines to accretion arcs on the surface of the white dwarf.  CVs are comprehensively discussed in both \citet{2003cvs..book.....W} and \citet{2001cvs..book.....H}, and see also \citet{2002apa..book.....F}.

Among the features seen in the photometry of some CVs are signals whose periods are a few percentage points larger and/or smaller than the binary's orbital period.  Collectively, these signals are called superhumps.  A particular signal is called a positive superhump if its period is greater than the orbital period.  Positive superhumps are attributed to the prograde precession of an oscillating eccentric disk as discovered by \citet{1988MNRAS.232...35W, 1988MNRAS.233..529W} and discussed in \citet{1991MNRAS.249...25W}, \citet[Chapter 5]{2001cvs..book.....H} or \citet[Section 3.6]{1995CAS....28.....W}.  Positive superhumps arise from two sources, as discussed in \citet{2001MNRAS.324..529R} and in detail in \citet{2011ApJ...741..105W}. The first source is periodic viscous dissipation in the oscillating  disk itself, and the second source is cyclical viscous dissipation as the accretion stream impact point sweeps around a still oscillating non-axisymmetric disk.  The first (disk) source dominates the light curve during superoutburst, and the second (stream or ``late'') source can dominate if the disk returns to quiescence while still oscillating.  

The so-called negative superhumps are less commonly observed.  They have a period slightly shorter than the orbital period and their origin is the subject of this paper.  Positive superhumps and negative superhumps may appear together or in isolation and the two seem to be independent.  Fewer systems are reported to have negative superhumps than positive superhumps: 21 are listed in the compilation provided in \citep{2009MNRAS.398.2110W} while 31 are listed in \citet{2009ApJ...705..603M}.  The 2012 version of the  catalog published by \citet{2011yCat....102018R} contains 31 systems though not all identifications are confirmed (see also Ritter \& Kolb 2003).  All in all, some forty-seven systems representing all classes of CVs except the polars may display negative superhump behavior.

There is general agreement that negative superhumps arise from the slow retrograde precession of the line-of-nodes of a tilted disk.  An early suggestion of this can be found in a study of TV Col \citep{1985A&A...143..313B} and in \citet{1993ApJS...86..235P}.  Analytical studies \citep{1995MNRAS.274..987P, 1998ApJ...497..212P, 1998ApJ...509..819T, 2000A&A...360.1031T} argue that such disks will precess in a retrograde direction but do not indicate how the disk may have become tilted nor do they discuss the source of the negative superhump signal.  Simulations using the method of smoothed particle hydrodynamics (SPH) that result in the retrograde precession of the disk and/or negative superhumps include \citet{1996MNRAS.282..597L}, \citet{2000ApJ...535L..39W}, \citet{2004PhDT........15M}, \citet{2009MNRAS.398.2110W}, and \citet{2009MNRAS.394.1897M} though in these simulations the disks are artificially tilted at some point during the simulation.  

Once we accept the possibility of a tilted disk, the source of the negative superhump signal can be understood.  This understanding comes from the recognition \citep{1988MNRAS.233..759B,1989MNRAS.236..735B, 2006MNRAS.366.1399F, 2007ApJ...661.1042W} that the accretion stream will impact a tilted disk at different radii during an orbit of the secondary.  Since the specific kinetic energy of the stream will increase the deeper into the gravitational potential the impact occurs, the luminosity of the bright spot will vary as it sweeps across first one face of the disk and then the other during a single orbit.   A distant observer sees only a single face of the disk and, hence, the slow retrograde precession of the disk yields an observed signal with a period slightly less than the orbital period -- the negative superhump period.  If we could monitor the total luminosity of the disk over time and integrated over 4$\pi$ sr, we would observe a signal at twice the nominal negative superhump frequency -- this is what our computer code yields as we discuss below.

As an analogy, consider a sine wave (period $T$ or frequency, $\omega_0 = 2\pi/T$) that is, in one case, half-wave rectified and in the second case undergoes full-wave rectification.  The Fourier series of the original, un-modified signal will have only one term in the amplitude spectrum and it will be at a frequency of $\omega_0$.  On the other hand, the Fourier Series representation of the half-wave rectified signal will contain frequency components at $\omega_0$, $2\omega_0$, $4\omega_0$, etc., while the full-wave rectified signal will have frequency components starting at $2\omega_0$ and continuing with $4\omega_0$, $6\omega_0$, etc., in the same manner as the half-rectified case.  The only difference between these two cases is in the location of the fundamental frequency--$\omega_0$ in one case and $2\omega_0$ in the other.  Returning to the case of negative superhumps, the observed negative superhump signal corresponds to the half-wave rectified sinusoid (we do not see the bright spot when it is on the 'back' side of the disk.)  The full-wave rectified signal corresponds to our observation of the total luminosity since, as the line-of-nodes rotates through 180$^\circ$, a new trace of the accretion stream across the face of disk replicating the previous sweep will begin.   

To complete the explanation of how negative superhump signals arise, we must determine the cause of the tilt.  A number of possible sources for disk warping and tilting have been examined including mis-alignment of the spins or angular momenta in the system \citep{1975ApJ...195L..65B, 1973NPhS..246...87K, 1972AZh....49..921S}, stream-disk interaction \citep{2009AcA....59..419S, 2012ApJ...745L..25M, 2012ApJ...753L..27M}, winds from the primary \citep{2001ApJ...563..313Q}, tidal effects due to the secondary \citep{1992ApJ...398..525L, 1996MNRAS.282..597L, 1998MNRAS.300..561M, 1998MNRAS.299L..32L}, irradiation from the primary \citep{1989Ap&SS.158..205H, 1996MNRAS.281..357P,1997ApJ...491L..43M, 1998MNRAS.300..561M, 1998ApJ...504...77M, 1996ApJ...472..582M, 1999MNRAS.308..207W, 2001MNRAS.320..485O}, radiation from an external source \citep{2008MNRAS.384..123I}, a magnetic field on the secondary \citep{2002MNRAS.335..247M}, and a magnetic field on the primary \citep{2004ApJ...604..766P, 1989Ap&SS.158..205H, 1999ApJ...524.1030L}.  

The results of these various studies are mixed.  A (purely) tidal approach to disk distortion has been shown to be too weak to be the either the source of tilt in CVs \citep{1998MNRAS.300..561M} or, if our premise is true, of negative superhumps.  Irradiation has been studied largely in the context of X-ray binaries.  In this case, irradiation has been shown to be a viable, if unlikely, source of warping and tilting in disks \citep{2001MNRAS.320..485O, 1998MNRAS.300..561M} though in \citet{2006MNRAS.366.1399F} SPH simulations have demonstrated the creation of both positive and negative superhumps in systems similar to X-ray binaries.  Winds have also been suggested as a source of warping both in X-ray binaries and in active galactic nuclei \citep{2001ApJ...563..313Q} though this concept has not been applied, apparently, to the generation of negative superhumps.  \citet{2009AcA....59..419S}, \citet{2012ApJ...745L..25M, 2012ApJ...753L..27M} and \citet{2010ApJ...722..989M} argue for a disk-stream interaction origin for negative superhumps, where the second author, \citeauthor{2012ApJ...745L..25M}, provides evidence obtained from the same basic simulation tool we use in this work.  In \citet{2002MNRAS.335..247M} simulations are reported that show a retrogradely precessing warped disk can arise in systems containing a magnetic field on the secondary.  

\citet{1999ApJ...524.1030L} has shown that precession and tilting can occur in disks when a magnetic field is present on the primary.  In \citeauthor{1999ApJ...524.1030L}'s \citeyear{1999ApJ...524.1030L} paper, the spin axis of the white dwarf is assumed to be offset from the rotation axis of the binary system.  A magnetic dipole field present on the white dwarf is, itself, tilted away from the spin axis of the white dwarf and, as a result, rotates about the spin axis at the spin rate.  \citet{1999ApJ...524.1030L} demonstrated, using the magnetic field structure discovered in \citet{1980A&A....86..192A}, that depending upon the way in which the magnetic field interacts with the disk, either tilting (the field lines thread the disk) or precession (the field lines are screened from the disk's interior) exists.  He proceeded to present a hybrid model that produced both behaviors.  In at least one system, the intermediate polar XY Ari \citep{2007A&A...472..225N}, just such a mechanism is used to explain the observed behavior of the system.  In \citeauthor{1999ApJ...524.1030L}'s \citeyear{1999ApJ...524.1030L} work, while the rate of tilt and the precession rate are developed, the possibility of a magnetic field origin for negative superhumps is not argued rather the main thrust of the paper is directed elsewhere.  Recently, \citet{2013ARep...57..327B} have also investigated the presence of a magnetic field on the white dwarf as a potential source of tilted disks in CVs.  Their work assumes a white dwarf that is synchronously locked to the orbital period of the binary with the magnetic dipole fixed in the co-rotating frame.  With the spin axis of the white dwarf aligned with the orbital motion, the tilting instability analyzed by \citet{1999ApJ...524.1030L} does not appear.  Their work, though different from \citeauthor{1999ApJ...524.1030L}'s \citeyear{1999ApJ...524.1030L} approach, demonstrates that tilted disk can arise but that the tilt eventually dampens out.  
  
In this paper we add, using the theory developed by \citet{1999ApJ...524.1030L}, the effects of a magnetic dipole located on a spinning primary to an existing SPH model originally written to model non-magnetic cataclysmic variable stars (see \citet{1995ApJ...448..822S} and \citet{1998ApJ...506..360S}).  Using this code, we demonstrate the emergence of negative superhumps in the simulation ``light curve'' over a range of magnetic dipole moments.  Except when the dipole moment is too small and no signal appears or when the dipole moment becomes sufficiently large that our simple model is no longer applicable and the period is not well defined, the period of the negative superhump is essentially fixed for a given mass ratio.    We show that the period deficits calculated from these negative superhumps match those calculated in simulations using manually tilted disks.  When positive superhumps appear, we also show that the period excesses calculated are also consistent with previous results.  Using examples, we show that the disk's are tilted, though the tilt varies in time, and precess in a retrograde manner.  The magnetic moments necessary for the creation of negative superhumps are estimated to vary over the range $10^{28}$-$10^{31}$ G cm$^3$.  Corresponding to kilogauss and weaker magnetic inductions, these fields are quite small and may explain why negative superhump systems are, with the exception of the IPs, considered non magnetic.  This result is consistent with the results provided by \citet{1998PASP..110..403P} where it is stated that magnetic fields below $10^5$ G are not easily detected using Zeeman line splitting techniques and that white dwarf stars with these fields may be common \citep[see also][]{warner2004}.  Ignoring the VY Scl stars, whose secondary may be magnetic and create negative superhumps via this route as discussed above, and those NLs whose magnetic nature has not yet been determined; the ratio of magnetic systems (the IPs and the SW Sex systems) to total systems (magnetic primaries and the DNs) is about 39 percent giving some credibility to a magnetic basis for this phenomena. 

At least three limitations are present in our approach.  First, our selection of only those simulations that show simple negative superhump behavior (i.e., a single spectral line) suggests that we are choosing planar disks.  This idea is reinforced given that our findings are identical to previous work by our group in which we manually tilted the disks and restarted the simulations \citep[e.g.,][]{2000ApJ...535L..39W,2007ApJ...661.1042W,2009MNRAS.398.2110W}.  In point of fact, our model is most accurate for planar disks since we use the disk's angular momentum as a proxy for the local normal to the disk.  Clearly for disks that are not planar (e.g., warped) this assumption is not warranted.  While we do not discuss here the results obtained in our simulations for high magnetic moment disks, deviations from planarity are seen, and the disk tilts are large enough that we believe our simple approach is no longer applicable, but rather that this regime of parameter space requires a more comprehensive approach using full magnetohydrodynamics.

\section{Approach}
\subsection{Basic Hydrodynamics}

We take as our starting point the Navier-Stokes equations \citep{1998pfp..book.....C} which describe the conservation of mass, momentum, and energy in a fluid.  In a form appropriate to our needs, these three conservation laws take the form:
\begin{eqnarray}
{d\rho \over {dt}} &=& -\rho {\nabla\cdot\bf{v}} \nonumber \\
{d^2{\bf r}\over{dt^2}}&=&-{\nabla P\over \rho}
+{{\bf f}_{\rm v}}
+{{\bf f}_{\rm m}}
+{{\bf f}_{\rm g}} \\
{{du}\over{dt}}&=&-{P\over\rho}{\nabla\cdot\bf{v}}+{\epsilon_{\rm v}} \nonumber ,
\end{eqnarray}

\noindent
where $\bf {f}_{\rm g}$ is the gravitational force (we use lower case $\bf {f}$ to signify force per unit mass), $\bf {f}_{\rm v}$ is the viscous force, and ${\bf f}_{\rm m}$ is the magnetic force and the subject of this paper, 
$\epsilon_{\rm v}$ is the energy generation from viscous dissipation, and the remaining symbols have their usual definitions.

Mass is automatically conserved in our version of SPH and we will not discuss this further.  The standard SPH energy equation is replaced in our code by an action-reaction scheme as explained in \citet{1998ApJ...506..360S} and described later.  All that remains is the momentum equation and it may be rewritten, by specifying the gravitational force, as 
\begin{eqnarray}
{d^2{\bf r}\over{dt^2}}=&-&{\nabla P\over \rho}+{{\bf f}_{\rm v}}+{{\bf f}_{\rm m}} \nonumber \\[1.4ex]
&-&GM_1{\bf{r - r_1}\over {\left|\bf{r - r_1} \right|^3}} 
- GM_2{\bf{r - r_2}\over {\left|\bf{r - r_2} \right|^3}},
\label{EqMom}
\end{eqnarray}

\noindent
where, $G$ is the gravitational constant, $M_1$ and $M_2$ are the masses of the primary and secondary, respectively, $\bf{r}$ is the distance vector from the system's center of mass to the point of interest, $\bf{r}_1$ is the vector from the center of mass to the primary, and $\bf{r}_2$ is the position vector of the secondary.  

In our code the equations are normalized to a non dimensional form, we transform $\bf{r}$ to $a\bf{R}$ and $\tau$ to $\Omega t$ where $a$ is the distance separating the two stars and $\Omega$ is the orbital frequency of the binary.  The parameters $a$ and $\Omega$ carry the dimensional information while $R$ and $\tau$ are dimensionless.  As a result of these transformations, accelerations ($\rm m\ s^{-2}$) in the physical domain are $\Omega ^2 a$ times their dimensionless simulation values.  Using Kepler's third law, we find that the momentum equation may now be written as ($q \equiv M_2 / M_1$):
\begin{eqnarray}
{d^2{\bf R}\over{d{\tau}^2}}=&-&{1 \over {\Omega ^2 a}} \left( {{\nabla P \over \rho}
-{{\bf f}_{\rm v}}
-{{\bf f}_{\rm m}}} \right) \nonumber \\[1.4ex]
&-& {1 \over {1+q}}{\bf{R - R_1}\over {\left|\bf{R - R_1} \right|^3}} 
- {q \over {1+q}}{\bf{R - R_2}\over {\left|\bf{R - R_2} \right|^3}}.
\label{EQ:SPHMom}
\end{eqnarray}

With the exception of the magnetic force ${{\bf f}_{\rm m}}$, this is the equation solved by the existing code.  We now discuss the form of the magnetic term ${{\bf f}_{\rm m}}$. 

\subsection{Forces on the disk due to the magnetic dipole} 

\begin{figure}
\plotone{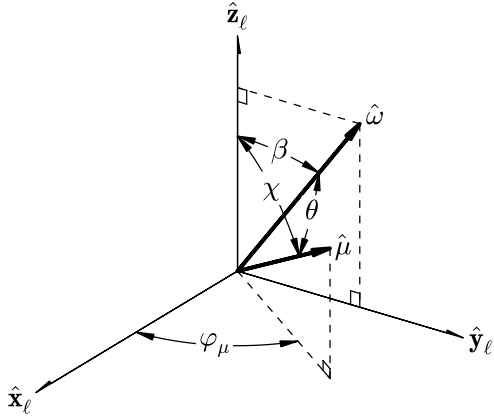}
\caption{The coordinate system used in calculating the magnetic forces on the disk is adapted from Figure 1 in \citep{1999ApJ...524.1030L}. The $z_\ell$-axis is aligned with the angular momentum of the disk ($\bl$) though the two are separated due to the presence of the secondary.  The angular velocity vector of the star ${\bo =\omega \bohat)}$ is inclined at an angle $\beta$ with respect to {$\bl$} and lies in the $y_{\ell}z_{\ell}$-plane. The stellar dipole moment, $\bmu$, rotates around the spin axis $\bo$ at an angle $\theta$, the angle of obliquity, and with $\varphi_{\mu}$ its instantaneous azimuth.  The time-dependent angle between the angular momentum vector and the magnetic moment vector is designated by $\chi$. (Note: All angles, except $\varphi_{\mu}$, are exaggerated for clarity.) 
\label{fig:f1}}
\end{figure}

Consider a magnetic dipole ($\bmu$) rotating about the spin axis ($\bo$) of a white dwarf as shown in Figure \ref{fig:f1}. The orientation of the spin axis is fixed in inertial space with known spin $\omega = \left| \bo \right|$.  The magnetic dipole, with strength $\mu = \left| \bmu \right|$, is tilted with a (known) fixed angle of $\theta$ relative to the spin axis.  The angular momentum of the disk is $\bl$ and is offset from the spin axis with an angle $\beta$.  As can be seen in the Figure, the coordinate system is chosen to have its $z$-axis aligned with the disk's angular momentum and so that the spin vector of the white dwarf is in the \textit{yz}-plane.

\citet{1999ApJ...524.1030L} has determined that the vertical force (per unit area) on the disk due to the magnetic dipole on the white dwarf may be written as (we note that Lai ignored, as do we, radial and azimuthal force terms arising from the white dwarf's magnetic field): 
\begin{equation}
F_{\ell}\blhat = -{\mu^2\over \pi {\left|\bf{r - r}_1 \right|}^6}\Gamma\left( {D,\zeta, {\omega}, \beta, \theta, \varphi } \right)\blhat, 
\label{forceLai}
\end{equation}
\noindent with
\begin{eqnarray}
\Gamma &=& {4\over\pi D}\sin\beta\sin\theta
\sin\omega t\sin\chi\cos(\varphi-\varphi_\mu) \nonumber \\[1.4ex]
&&+ {\zeta\over 2}\cos\beta\cos\theta\sin\chi\sin(\varphi-\varphi_\mu),
\label{EqGamma}
\end{eqnarray}
\noindent
and where $\bf{r}$ is measured in a cylindrical coordinate system.  The time dependent angles $\varphi_{\mu}$ and $\chi$ are defined by the following relations: 
\begin{eqnarray}
\sin\chi\cos\varphi_\mu & = & \sin\theta\cos\omega t, \label{id1} \nonumber \\
\sin\chi\sin\varphi_\mu & = & \sin\beta\cos\theta+\cos\beta\sin\theta\sin\omega t,
 \label{id2}  \\
\cos\chi & = & \cos\beta\cos\theta-\sin\beta\sin\theta\sin\omega t . \nonumber
\label{id3}
\end{eqnarray}
Because of these relationships, $\varphi_\mu$ and $\chi$ can be eliminated from the definition of $\Gamma$ given in Equation~\ref{EqGamma} and used in Equation~\ref{forceLai}.  Doing this and grouping the results in a slightly different way gives
\begin{eqnarray}
F_{\ell} = 
&-&{{\mu^2} \over {\pi^2D r^6}} \sin{2\beta}\sin^2\theta \sin\varphi \nonumber \\
&-&{{2\mu^2} \over {\pi^2D r^6}} \sin^2\beta \sin{2\theta} \sin\varphi \sin\omega t \nonumber \\
&-&{{2\mu^2} \over {\pi^2D r^6}} \sin\beta \sin^2\theta \cos\varphi  \sin{2\omega t} \nonumber \\
&+&{{\mu^2} \over {\pi^2D r^6}} \sin{2\beta}\sin^2\theta \sin\varphi\cos{2\omega t} \nonumber \\
&-&\zeta {{\mu^2} \over {4\pi r^6}}\cos\beta \sin{2\theta} \sin\varphi \cos\omega t \nonumber \\
&+&\zeta {{\mu^2} \over {4\pi r^6}}\cos^2\beta \sin{2\theta} \cos\varphi \sin\omega t  \nonumber \\
&+&\zeta {{\mu^2} \over {4\pi r^6}}\sin{2\beta} \cos^2\theta \cos\varphi.
\label{EQ:FourierForce}
\end{eqnarray}
While the majority of results cited in this paper use the full (again, ignoring certain radial and azimuthal terms) force model, in the small number of cases that we have investigated, only the first term in Equation~\ref{EQ:FourierForce} is needed to the produce same result as the full model (see Lai 1999 for the expression for the magnetic torque).  Additional information regarding the parameters of this model can be found in Table \ref{tLai}.  

As we are interested in this work with the negative superhump phenomena, we have not investigated the effects of terms 2-5 in the expression for $F_{\ell}$ given above.  For a discussion of the effects of these terms on disk dynamics see \citet{2008ApJ...683..949L}.

\begin{deluxetable*}{ccl} 
\tablecolumns{3} 
\tablewidth{0pt} 
\tablecaption{Magnetic Dipole Model Parameter Description} 
\tablehead{\colhead{Symbol} & \colhead{Units}  & \colhead{Description}}
\startdata 
$\beta$ & radians & Angle between $\bl$ and $\bo$ \\
$\chi$ & radians & Angle between $\bl$ and $\bmu$ \\
$D$ & \nodata & $D=\max \left({\sqrt{\left( r/r_m \right) ^2 - 1}, \sqrt{2H/r_m}} \right)$ \\
$H$ & cm & Vertically measured disk half-height \\
$\bl$ & g cm$^2$ s$^{-1}$ & Angular momentum of disk \\
$\ell$ & g cm$^2$ s$^{-1}$ & Angular momentum of disk $\left( \ell = \left|{\bl} \right| \right)$ \\
$\bmu$ & Gauss cm$^3$ & White dwarf magnetic dipole moment \\
$\mu$ & Gauss cm$^3$ & Magnetic dipole moment strength $\left( \mu = \left|{\bmu} \right| \right)$ \\
$\bo$ & $s^{-1}$ & White dwarf spin vector \\
$\omega$ & $s^{-1}$ & White dwarf spin $\left( \omega = \left|{\bo} \right| \right)$ \\
$\varphi$ & radians & Angle of location in $x_{\ell}$$y_{\ell}$-plane \\
$\varphi_\mu$ & radians & Azimuth of $\bmu$ in $x_{\ell}$$y_{\ell}$-plane \\
$r$ & cm & Distance from center of white dwarf \\
$r_m$ & cm & Location of magnetospheric boundary \\
$\theta$ & radians & Time varying angle between $\bo$ and $\bmu$ \\
$\zeta$ & \nodata & Azimuthal pitch of the magnetic field lines
\enddata 
\label{tLai}
\end{deluxetable*}

\section{Simulation Approach}
\subsection{Smoothed Particle Hydrodynamics}

We model accretion disk dynamics using an SPH code first described in \citet{1995PhDT........11S, 1995ApJ...448..822S}.  Additional discussion of this code may be found in \citet{1998ApJ...506..360S}, \citet{2007ApJ...661.1042W}, and \citet{2006PASP..118..442W}. General treatments of SPH may be found in 
\citet{1990nmns.work..269B}, 
\citet{1992ARA&A..30..543M, 2005RPPh...68.1703M},
\citet{2009NewAR..53...78R}, 
and \citet{2010ARA&A..48..391S}.
The relationship between the Navier-Stokes Equations and SPH as applied to accretion disk dynamics is presented in \citet{1995CoPhC..89....1R}.       

The SPH form of the momentum equation (See Equation \ref{EqMom}) for the $i^{th}$ particle is
	\begin{eqnarray} 
	{d^2{\bf r}_i\over{dt^2}}
	&=& -\sum_j m_j\left({P_i\over{{\rho_i}^2}}	+{P_j\over{{\rho_j}^2}}\right)\left(1+\Pi_{ij}\right){\nabla_i{W_{ij}}}\nonumber\\[1.4ex]
		&&- GM_2{\bf{r - r_1} \over {\left|\bf{r - r_1} \right|^3}}
		  - GM_1{\bf{r - r_2} \over {\left|\bf{r - r_2} \right|^3}} + {\bf f}_{\rm m},
	\label{sphforce}
\end{eqnarray}
where $m_j$ is the mass of particle $j$ (we assume a constant smoothing length and equal mass for all particles) and $W_{ij}$ is the SPH kernel function and where the reduction to dimensionless form is yet to be done.
We use the artificial viscosity approximation of \citet{lattanzio86} evaluated between particles $i$ and $j$,
\begin{equation}
\Pi_{ij} = \left\{ \begin{array}{ll} 
		-\alpha\mu_{ij} + \beta\mu_{ij}^2 & {\bf v}_{ij}\cdot {\bf r}_{ij} \le 0;\\[1.4ex]
		0  & \mbox{otherwise;}
		\end{array}\right. 
\end{equation}
where
\begin{equation}
\mu_{ij} = {h{\bf v}_{ij}\cdot {\bf r}_{ij} \over c_{s,ij}(r_{ij}^2 + \eta^2)},
\end{equation}
and $\bf v _{ij} = \bf v _i - \bf v _j$, $\bf r_{ij} = \bf r_i - \bf r_j$.  For the simulations presented here, we set $\alpha=1.0$, $\beta=0.5$,  and $\eta=0.1h$ ($h$ is the SPH smoothing length).  The sound speed of particle $i$ is given by $c_{s,i} = \sqrt{\gamma(\gamma-1)u_i}$.  The average of the sound speeds for particles $i$ and $j$ is $c_{s,ij}$.  We use a gamma-law ideal gas equation of state $P=(\gamma-1)\rho u$ with $\gamma=1.01$.

In any given simulation, a fixed smoothing length is used whose value is determined, essentially, by the number of particles in the simulation with the goal of establishing a reasonable number of nearest neighbors in the smoothing process.  Each SPH particle has its own time step varied, as necessary, to resolve the local dynamics.  The largest timestep a particle can have is $\Delta t^0=\Porb/200$ where $\Porb$ is the orbital period of the CV.  Other timesteps available are $\Delta t^k = \Delta t^0/2^k$, where $1 \le k \le k_{\rm max}$ with $k_{\rm max}=10$.  We build up our model disks by injecting particles in a small region near $L_1$ at a rate of 2000 particles per orbit until the desired maximum number of particles in the disk is reached (25,000 for the simulations presented here).  Any time that a particle passes within $0.03a$ of the primary, or the effective radius of the secondary, or is ejected from the system, a replacement particle is injected at the $L_1$ point.  Thus the disk keeps a constant number of particles once the target number is reached.  Note that we do not model the boundary layer in this work

As discussed in \citet{1998ApJ...506..360S}, internal energies are propagated using an action-reaction principle in a process that those authors demonstrated to be equivalent to the standard SPH energy equation.  At any given time, we represent the change in bolometric luminosity by the sum of the changes in the internal energy over all the particles during the previous time step.  With this approach we can estimate a simulation ``light curve'' as a time series with time index $n$ as 
\begin{eqnarray}
L\left( n \right) \equiv L\left( n \Delta t_0 \right) & = & \sum_j {{d} \over dt} u_j\left( n \Delta t_0 \right). 
\label{eq:lc}
\end{eqnarray}
It is the frequencies present in these light curves that we examine for superhump activity.
\subsection{Incorporating a Magnetic Dipole into SPH}
The magnetic force given in Equation~\ref{forceLai} is specified as a force per unit area.  This suggests that acceleration due to the magnetic field can be expressed as 
\begin{equation}
{\ddot {z}_\ell} = f_m = {{F_{\ell}} \over {\Sigma}}
\label{EQ:ForceOArea}
\end{equation}
where $\Sigma$ is the vertically integrated density.  This integrated density is not readily available in our code and so we need to rewrite the equation above in terms of the more accessible density $\rho$.

Even with this change, we still face a problem.  As we have discussed previously, distances in our code are scaled by the separation between the two stars, $a$, while time is scaled by the orbital frequency $\Omega$.  These scalings are conventional and are dealt with easily. Dealing with mass and mass related entities is not so simple.  Since each particle is assumed to have mass one and since the accretion rate is not modeled (in fact, we assume that the system has achieved steady state and mass is added only to maintain the steady state), the relation between terms containing mass in the simulation and those representing the equivalent physical value is not well defined. 
 
The approach we take here is to assume that the (simulated) mass of the disk is equal to the number of particles used in the simulation (since each particle has mass equal to unity).  We then consider the following relations (a subscript of 's' refers to the simulation or scaled value while an unsubscripted parameter refers to the equivalent physical value. The exceptions to this rule, $F_{\ell}$ and $\ddot {z}_\ell$, are both physical values.)
\begin{eqnarray}
\ddot {z}_\ell & = & \frac{F_{\ell}}{\Sigma} = \frac{F_{\ell}}{\rho H} \nonumber\\
& = & \frac{F_{\ell}}{{ \left( {\rho_{s}}/{\rho_{s}} \right) \rho H}} = \frac{F_{\ell}}{{ \left( {\rho}/{\rho_{s}} \right) H \rho_{s}}} \nonumber\\
& = & \frac{F_{\ell}}{ \left( {M}/{V} \right) \left( {V_{s}}/{M_{s}} \right) H \rho_{s}} \\
& = & \frac{F_{\ell}}{ \left( {M}/{M_{s}} \right) \left( {V_{s}}/{V} \right) H \rho_{s}} \nonumber\\
& = & \frac{F_{\ell}}{ \left( M/N \right) \left( {V_{s}}/{a^3 V_{s}} \right) H \rho_{s}}. \nonumber
\end{eqnarray}
The second equality given above can be found in \citet[Equation 5.41]{2002apa..book.....F} while the remaining steps are straightforward.

We are left with the following relationship:
\begin{equation}
{\bf f}_{\rm m} = \ddot {z}_\ell = {\left( \frac{N a^3}{ M H} \right)} \frac{F_{\ell}}{\rho_{s}}.
\end{equation}
Using the relation $\ddot {z}_\ell = \Omega^2a \ddot {z}_{\ell,s}$ allows us to write
\begin{equation}
\left({\bf f}_{\rm m}\right)_s = \ddot {z}_{\ell,s} =  {\left( \frac{N a^2}{ \Omega^2 M H} \right)} \frac{F_{\ell}}{\rho_{s}}.
\end{equation}
For the mass of the disk and it half-height, we take the Shakura-Sunyaev steady-state solutions (pressure dominated with free-free opacity) given in \citep[Equations 5.49 and 5.51]{2002apa..book.....F} modified so that distances are scaled by the stellar separation $a$.  This change of scaling is designated by the change from $R_{10}$ in the published equations to $R_a$ in the current work.  Leading constants also have been changed to reflect this re-scaling.  The relevant equations are given below while the definition of each parameter is given in Table \ref{tSS}.
\begin{equation}
M = M_{disk} = \left( {{{10}^{ - 10}}{M_ \odot }} \right) {\alpha_{SS}^{ - 0.8}} \dot M_{16}^{0.7},
\end{equation}
and,
\begin{equation}
H = {k_H} a^{1.125} \alpha_{SS}^{-0.1} \dot M_{16}^{0.15} m_1^{-0.375} {R}_a^{1.125} f^{0.6},
\end{equation}
where $k_H = 9.5598 \times {10^{-4}}$ cm$^{-0.125}$, $R_a$ is the distance from the primary, and
\begin{eqnarray}
f(\bf{r}) &=& \left(1 - \sqrt{ \frac{r_{\rm wd}}{\left|\bf{r - r_1} \right|}} \right)^{0.25} \nonumber\\
&=& \left(1 - \sqrt{ \frac{R_{\rm wd}}{\left|\bf{R - R_1} \right|}} \right)^{0.25} = f(\bf{R})
\label{EQ:f}
\end{eqnarray}
with $r_{\rm wd}$ the radius of the white dwarf primary.  

Substituting for $F_{\ell}$, $M$, and $H$ gives the acceleration on the particle due to the magnetic field
\begin{equation}
\ddot {z}_{\ell,s} =  - \frac{1}{{\pi {\rho _{s}}}}\frac{\mu_s^2}{{\left|\bf{R - R}_1 \right|}^{7.125}f^{0.6}(\bf{R})}\Gamma 
\label{EqAcc}
\end{equation}
where $\mu_s$, the {\it scaled dipole moment}, is defined by
\begin{equation}
\mu_s^2 = \frac{N \mu ^2}{{k_{\mu}^2 \Omega^2 a^{5.125} {\alpha_{SS}^{-0.9}} \dot M_{16}^{0.85} m_1^{-0.375}}},
\label{EqMuSq} 
\end{equation}
with $k_{\mu}^2 = 1.9 \times 10^{20}$ g cm$^{-0.125}$.  Equation (\ref{EqAcc}), using the definition of $\Gamma$ given in Equation (\ref{EqGamma}), is the basis for the results reported in this paper and is applied only to particles outside the Alfv\'{e}n radius (see Table~\ref{tcvDisk}).

The units of the scaled dipole moment $\mu_s$ are easily determined.  First, we note that $\mu$ has units of G cm$^{3}$ which may be rewritten as (g$^{0.5}$cm$^{-0.5}$s$^{-1}$)$\times$cm$^3$ \citep[See Table 3.1]{2001deec.book.....C}.  The orbital frequency $\Omega$ has units of s$^{-1}$ while $a$ is measured in cm.  The units of $k_{\mu}^2$ are g cm$^{-0.125}$. The other parameters are unitless.  A dimensional analysis yields,
\begin{equation}
\mu_s^2 \sim \frac {\left(\mbox{g cm}^{-1} \mbox{ s}^{-2} \right) \mbox{cm}^6}{\mbox{g cm}^{-0.125} \mbox{ s}^{-2} \mbox{ cm}^{5.125}} \nonumber
\end{equation}
and $\mu_s$ is found to be unitless.

While $\mu_s$ appears naturally in this approach, we find in our simulations that the scaled dipole moment has a value on the order of $10^{-4}$.  In order to keep our numbers on the order of unity, we introduce the notation $\mu_r \equiv 10^{4}\mu_s$ and $k_{r} \equiv 10^{-4}k_{\mu}$.

Finally, we remind the reader that two different radius vectors are used.  For terms involving the magnetic field force, the radius vector is expressed in a cylindrical coordinate system while all other relationships use a radius vector appropriate to a spherical coordinate system.

\begin{deluxetable*}{ccl} 
\tablecolumns{3} 
\tablewidth{0pt} 
\tablecaption{Shakura-Sunyaev $\alpha$-Disk Parameter Description} 
\tablehead{\colhead{Symbol} & \colhead{Units}  & \colhead{Description}}
\startdata 
$a$ & cm & Stellar separation \\
$\alpha_{SS}$ & \nodata & Shakura-Sunyaev viscosity parameter \\
$H$ & cm & Vertically measured disk half-height \\
$M_\odot$ & g & Solar mass $\left( 1.989 \times 10^{33} \right)$ \\
${M_{disk}}$ & g & Accretion disk mass \\
$m_1$ & \nodata & Primary mass relative to $M_\odot$ \\
$\dot M$ & $ g\ s^{-1}$ & Accretion rate \\
${\dot M_{16}}$ & \nodata & Accretion rate ${{\dot M} \mathord{\left/
 {\vphantom {{\dot M} {\left( {{{10}^{16}}\ g {s^{ - 1}}} \right)}}} \right.
 \kern-\nulldelimiterspace} {\left( {{{10}^{16}}\ g {s^{ - 1}}} \right)}}$ \\
$r$ & cm & Distance from center of white dwarf \\
$r_{wd}$ & cm & White dwarf radius \\
$R_a$ & \nodata & Distance $r$ normalized by $a$ \\
$R_\odot$ & cm & Solar radius $\left( 6.955 \times 10^{10} \right)$ \\
$\Sigma$ & $\ g\ cm^{-2}$ & Vertically integrated density
\enddata 
\label{tSS}
\end{deluxetable*}

\subsection{Coordinate systems and transformations}
In the code, we must transform back and forth between the system in which the magnetic forces are calculated and the system where these same forces are applied.  Forces are applied in a non-rotating system whose origin is the system's center of mass and where the \textit{z}-axis is perpendicular to the orbital plane of the two stars.  An intermediate system is centered on the white dwarf and obtained by simple translation of the center-of-mass system to the white dwarf.  The unit vectors describing this system will be designated, respectively, $\bxhat$, $\byhat$ and $\bzhat$.  The final system, also centered on the white dwarf, has its \textit{z}-axis aligned with the disk's angular momentum (unit vector $\blhat$) and is oriented so that the spin axis of the white dwarf (unit vector $\bohat$) lies in its \textit{yz}-plane.  Quantities measured in this system will have the subscript ${\ell}$.  For example, the unit vectors defining this system be designated $\bf{\hat x_{\ell}}$, $\bf{\hat y_{\ell}}$ and $\bf{\hat z_{\ell}}$.   This coordinate system, illustrated in Figure~\ref{fig:f1}, is the same as that used by \citet{1999ApJ...524.1030L} and is the system in which the magnetic forces, $\mathbf {f}_m = \ddot {z}_{\ell} \blhat$ or $\left({\bf f}_{\rm m}\right)_s = \ddot {z}_{\ell,s} \blhat$ in the scaled system, are calculated.  We can write the unit vectors in this system as (where $\blhat = {{\ell}_x}{\bf{\hat x}} + {{\ell}_y}{\bf{\hat y}} + {{\ell}_z}{\bf{\hat z}}$ and $\bohat = {\hat \omega _{x}}{\bf{\hat x}} + { \hat \omega _{y}}{\bf{\hat y}} + {\hat \omega _{z}}{\bf{\hat z}}$)
\begin{equation}
{\bxhat_\ell \equiv \frac{\bohat \times \blhat}{\left| \bohat \times \blhat \right|}},
\end{equation}
\begin{equation}
\byhat_\ell \equiv \bzhat_\ell \times \bxhat_\ell = \frac{{{\bohat}} - \left( {{{\bohat}} \cdot \blhat} \right) \blhat}{{\left| {\bohat - \left( {\bohat \cdot \blhat} \right) \blhat} \right|}},
\end{equation}
and,
\begin{equation}
\bzhat_\ell  \equiv \blhat.
\end{equation}
\noindent It is easy to demonstrate that $\left|{\bohat - \left( {\bohat \cdot \blhat} \right) \blhat}\right| = \left|\bohat \times \blhat \right|$, that $\bohat \times \blhat = \sin\beta \bxhat_\ell $, and that $\bohat \cdot \blhat = \cos\beta$.

We calculate $\blhat$ within the code as necessary while $\bo$ is given by the problem definition.  As a result, the coordinate system is well defined and routine coordinate transforms allow us to move between the different systems.  Care must be taken when $\bl$ is too close to $\bo$ since $\bxhat_\ell$ will be near zero in magnitude, an undesirable event.

\section{SPH Simulation Results}
\subsection{Tilted disks, precession, and superhumps}
\subsubsection{Approach}
Our modified SPH code was exercised for a variety of mass ratios ($q=$ 0.10, 0.15, $\ldots$ , 0.60) and scaled dipole moments ($\mu_r =$ 0.2, 0.4, $\ldots$ , 6.0) while holding all other parameters at a fixed value.  In each simulation 1,000 orbits, containing 25,000 particles, were computed.  Once this baseline was established, we allowed other parameters to vary in an effort to determine their effect.  By no means has the entire parameter space of the new model been exercised.  Still, we see the emergence of negative superhumps and tilting of the disks.  The parameter space is presented in Table \ref{tcvDisk}.  For comparison purposes, simulations were done for the same values of the mass ratio as above but without a magnetic field.  In the following, we demonstrate the emergence of negative superhumps, the creation of tilted disks, and the retrograde precession of these same disks.  

\begin{figure*}
\plotone{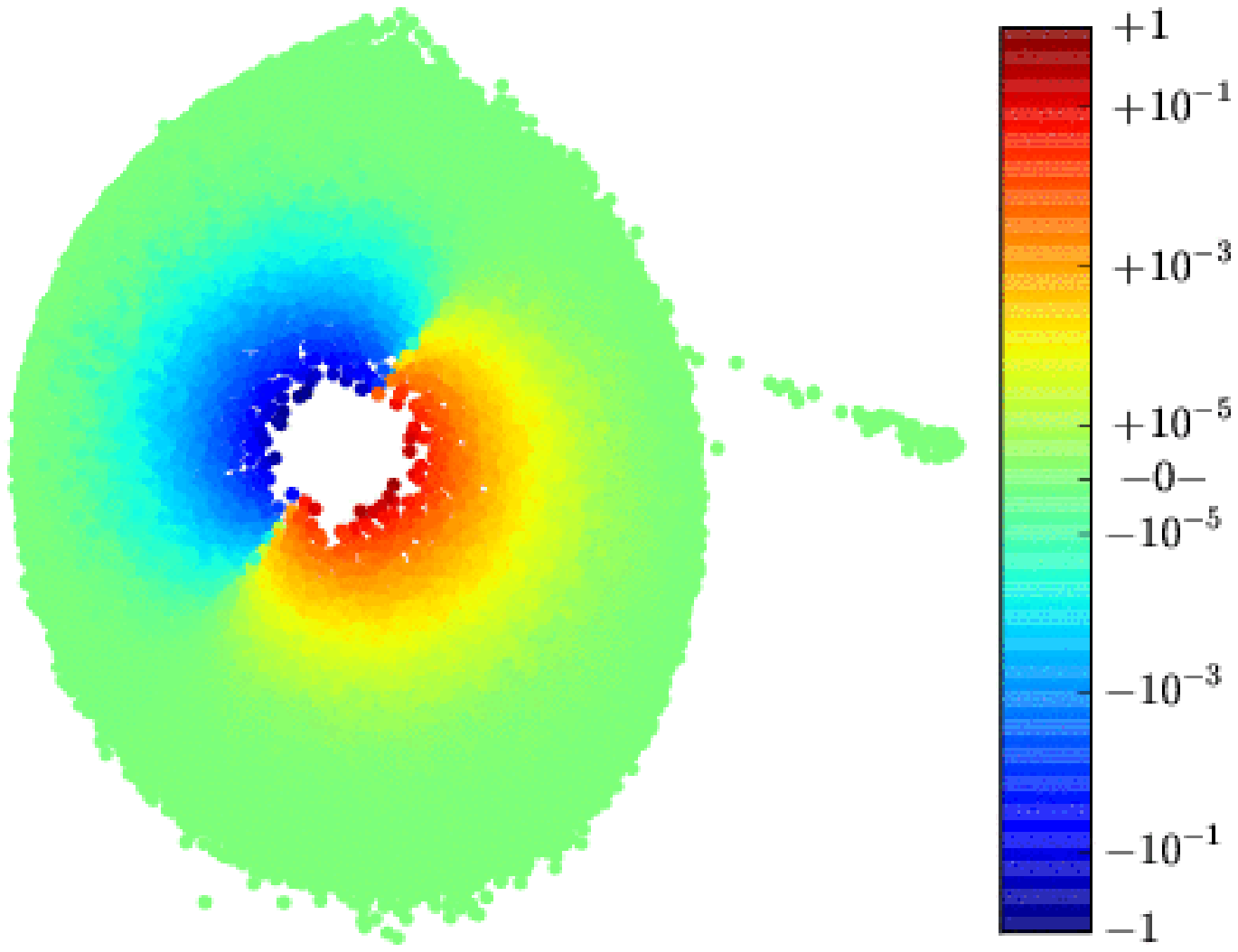}
\caption{Particle accelerations calculated within the simulation for orbit 724 with $q=0.3$ and $\mu_r = 3.0$ using Equation~\ref{EqAcc}.  The logarithm of the magnitude of the acceleration is shown using the color mapping, and the sign of the value indicates the direction of the acceleration.  The bipolar nature of the accelerations are clearly seen as is the sharp decay of the magnitude of the accelerations as the distance from the white dwarf primary is increased.  The disk is tilted at an angle of 2.08$^{\circ}$.  The disk is oriented so that the angular momentum vector of the disk points out of the page.  With this orientation, the magnetic field induced accelerations either point out of the page (aligned with the angular momentum) or point into the page (anti-aligned).  The simulated accelerations vary over the range -0.789 to +0.431 in system units, corresponding to physical accelerations of $-309.43$ to 169.22 m s$^{-2}$.  It should be noted that the scaling used in this illustration is logarithmic and that accelerations with magnitude less than 10$^{-5}$ are set to be equal to zero for display purposes.}
\label{fig:f2}
\end{figure*}

In Figure~\ref{fig:f2}, we show a view of a disk oriented so that the angular momentum vector of the disk is pointed out of the page.  The color mapping indicates the logarithm of the magnitude of the accelerations given by Equation 19, and the sign indicates the direction. Note that as expected these accelerations are bipolar and the magnitude of the magnetically induced acceleration experienced by disk particles rapidly decays as the distance from the white dwarf increases. 

\begin{deluxetable*}{cccl} 
\tablecolumns{4} 
\tablewidth{0pt} 
\tablecaption{Simulation Parameters} 
\tablehead{\colhead{Symbol} & \colhead{Value} & \colhead{Units}  & \colhead{Description}}
\startdata 
$\alpha$ & 1.0 & \nodata & $1^{st}$ Artificial viscosity parameter \\
$\beta$ & 0.5 & \nodata & $2^{nd}$ Artificial viscosity parameter \\
$r_m$ & $0.06a$ & cm & Alfv\'{e}n radius \\
$H/r$ & 3.5 & $\deg$ & Disk opening angle  \\
$H$ & Calculated & cm & Disk half height: $H = \left( H / r \right) r$ \\
$N$ & 25,000 & \nodata & Number of particles in simulation \\
$\Omega$ & Calculated & s$^{-1}$ & Binary orbital frequency\\
$\omega$ & 25$\Omega$ & s$^{-1}$ & Spin frequency\\
$\beta_\omega$ & 5 & $\deg$ & Spin vector tilt in inertial space\\
$\phi_\omega$ & 90 & $\deg$ & Spin vector azimuth in inertial space\\
$\theta$ & 1 & $\deg$ & Angle of obliquity \\
$\mu_s$ & Variable & \nodata & Scaled dipole moment strength \\
$\mu_r$ & Variable & \nodata & $10^4\mu_s$ \\
$m_1$ & 0.6 & \nodata & Primary mass (multiples of $M_\odot$) \\
$m_2$ & Variable & \nodata & Secondary mass (multiples of $M_\odot$) \\
$M_1$ & Variable & g & Primary mass\\
$M_2$ & Variable & g & Secondary mass\\
$m_1$ & 0.6 & \nodata & Primary mass (multiples of $M_\odot$)\\
$m_2$ & Variable & \nodata & Secondary mass (multiples of $M_\odot$)\\
$q$ & Calculated & \nodata & Mass ratio $m_2 / m_1$
\enddata 
\label{tcvDisk}
\end{deluxetable*} 

\subsubsection{Power Spectrum Analysis}

Once each simulation was complete, a Fourier analysis was performed.  For visualizations, the raw amplitude spectra, calculated using a fast Fourier transform (FFT) provided in the Python numerical library ``numpy'' and consisting of 100,000 points taken from the last 500 orbits (200 points per orbit) of the simulation, were used.  Subsequently, the precession frequency, if it appeared, was identified as were the frequencies of existing positive and negative superhumps.  For this purpose, we use a discrete Fourier transform that is oversampled in frequency in a region surrounding the possible peak together with peak interpolation to more precisely locate the frequencies.  Evaluation of the spectra provided shows that the identified peaks easily exceed a signal-to-noise ratio of 4, an often-used threshold criterion presented for observational time-series data in \cite{1993A&A...271..482B}.  

\begin{figure*}
\plottwo{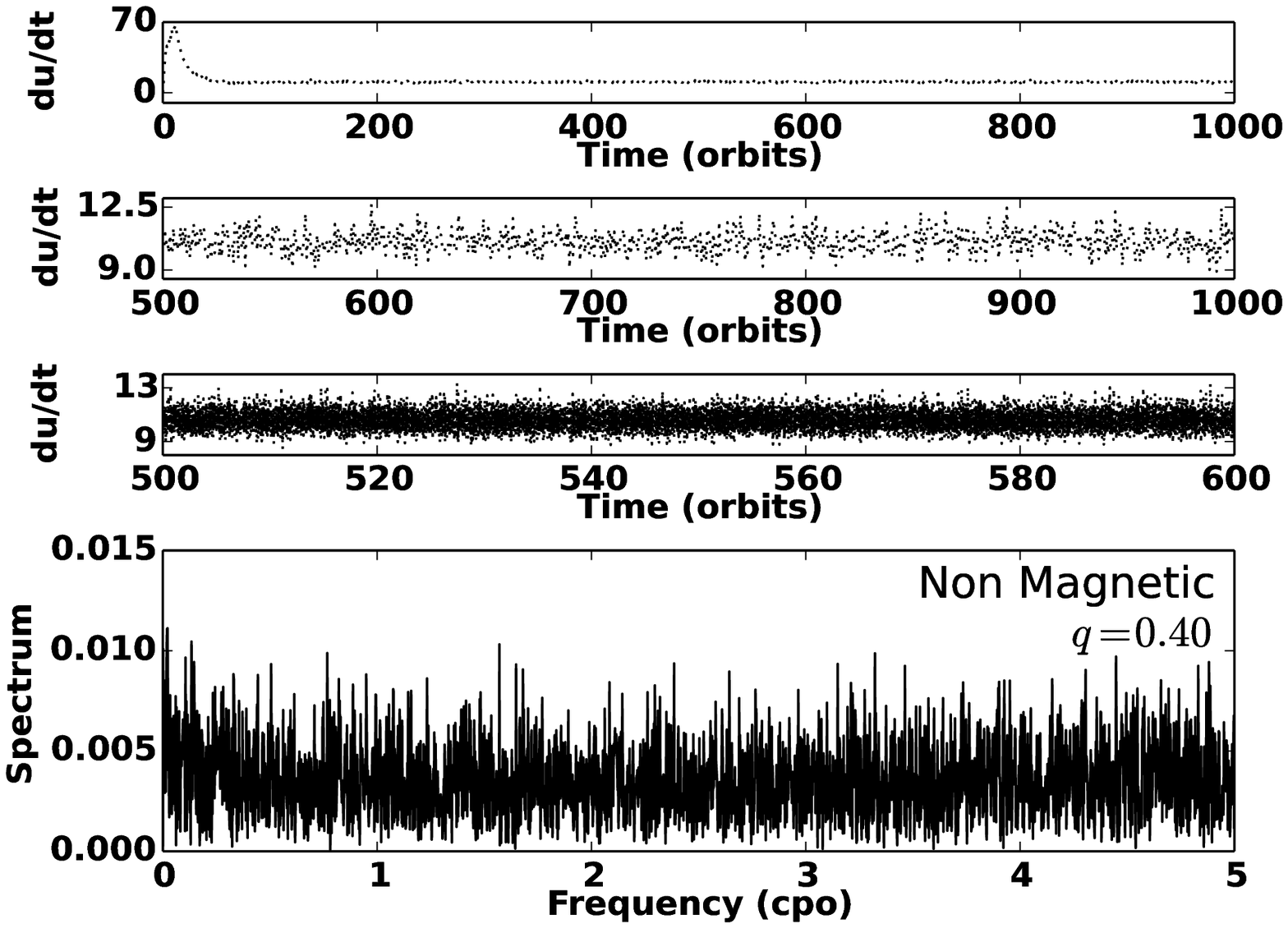}{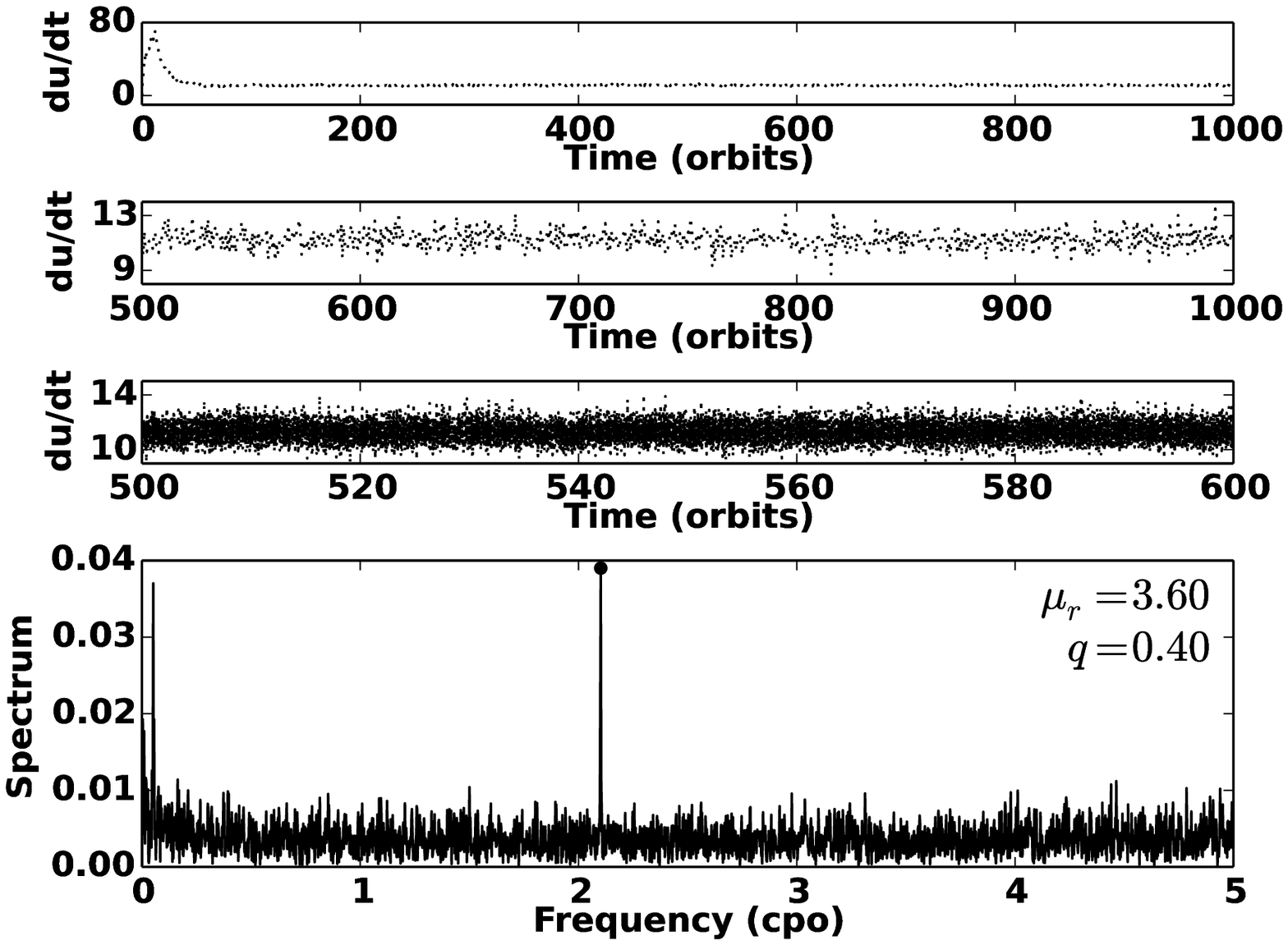}
\caption{Magnetically Induced Negative Superhump: Left column is a Non Magnetic System with q = 0.4: (Top Three Plots) Selected segments of the time series, (Bottom Plot) Amplitude Spectrum. Right Column: Same as left with $ \mu_r = 3.6$.  Note the appearance of a negative superhump (identified by the filled circle) in the spectrum on the right.  Frequency is measured in cycles per orbit or cpo.  See the text for discussion of the processing used to prepare these and the following spectra for display.}
\label{fig:f3}
\end{figure*}

As our first example, we consider a system with a mass ratio of $q=0.4$ and compare the amplitude spectrum of the simulated light curve for a non-magnetic system and a magnetic system with a scaled dipole moment equal to $ \mu_r = 3.6$.  In Figure~\ref{fig:f3} the two cases are presented in, respectively, the left and right columns.  The top graph in each column shows selected segments of the simulated light curve.  For visualization purposes, the top graph is sampled every 110 points, the middle plot is sampled every 50 points, and the bottom light curve is sampled every point.  The bottom graph in each column is the amplitude spectrum calculated using an FFT as described earlier.  As expected, the non-magnetic system contains neither positive superhumps (the mass ratio is too large) nor negative superhumps.  In the magnetic system, we do see a negative superhump signal at $\nu_- = 2.1$ cycles per orbit (cpo) as indicated by the filled circle identifying the spectral peak.

\begin{figure*}
\plottwo{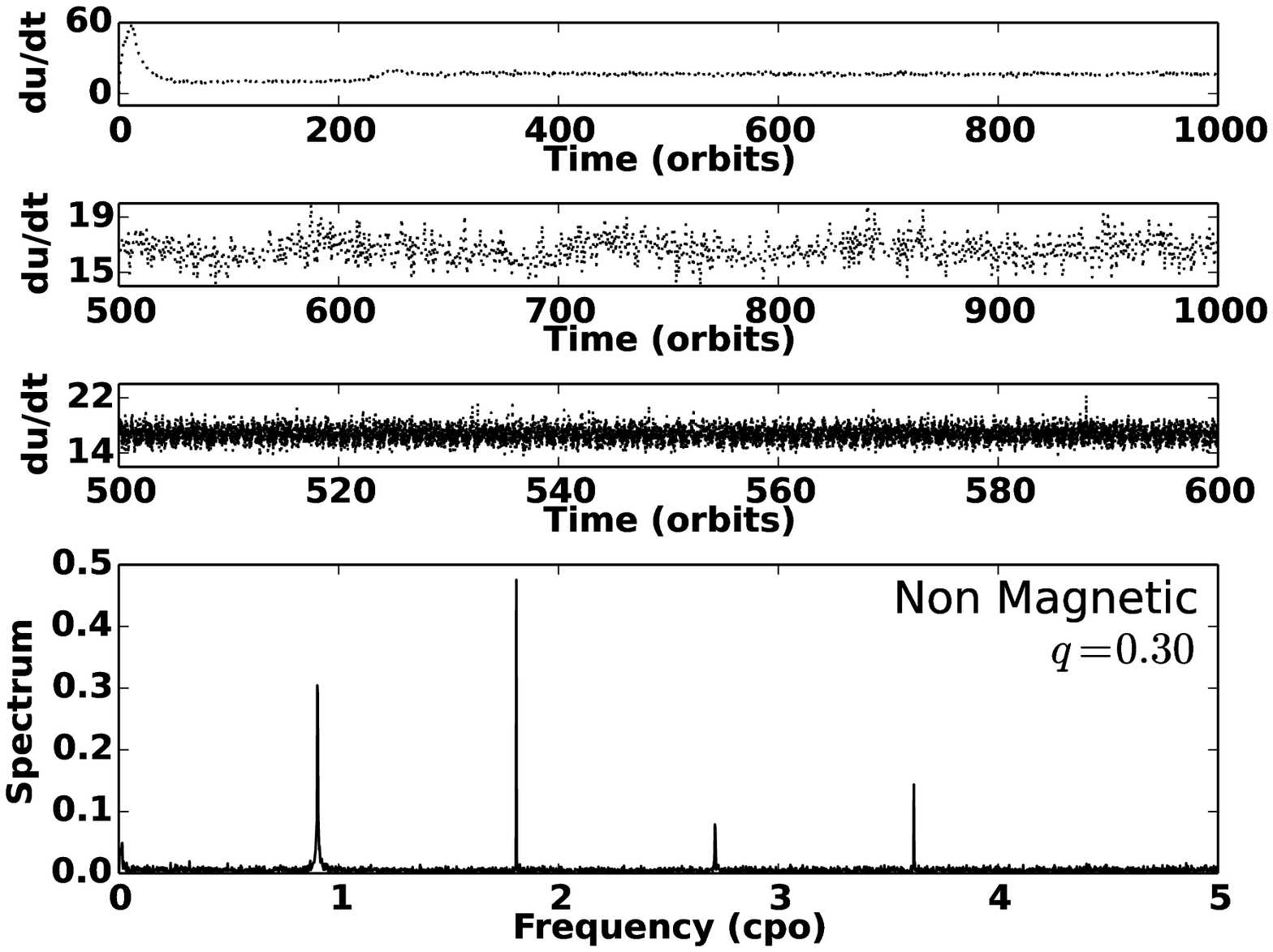}{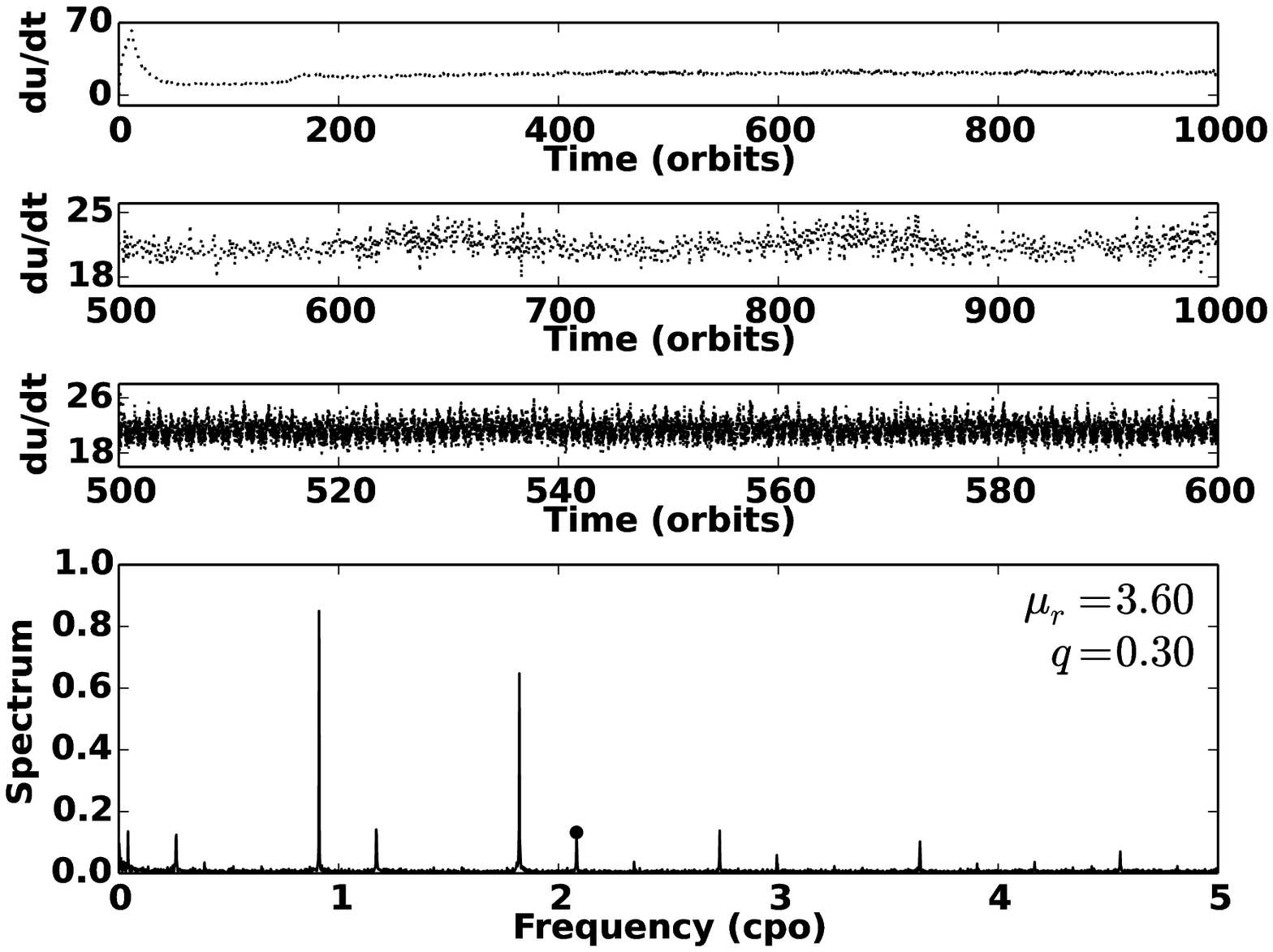}
\caption{Magnetically Induced Negative Superhump: Left column is a Non Magnetic System with q = 0.3:  (Top Three Plots) Selected segments of the time series, (Bottom Plot) Amplitude Spectrum. Right Column: Same as left with $\mu_r = 3.6$.  Note the appearance of a negative superhump (identified by the filled circle) in the right spectrum.}
\label{fig:f4}
\end{figure*}

We next consider a system expected to possess positive superhumps.  In this example, the same in all respects as in the previous example save that the mass ratio $q$ is set to the value $0.3$, a positive superhump signal is seen in the non magnetic system as shown on the left hand side of Figure~\ref{fig:f4}.  We also see a number of higher harmonics of the positive superhump fundamental frequency ($\nu_+$).  In the right column, we see that the same positive superhump signal(s) appears but we now also see negative superhumps (indicated by the filled circles).  Additional frequencies can appear at sum and difference combinations of the positive and negative superhump frequencies, $\nu_+$ and $\nu_-$, respectively.

\begin{figure*}
\plottwo{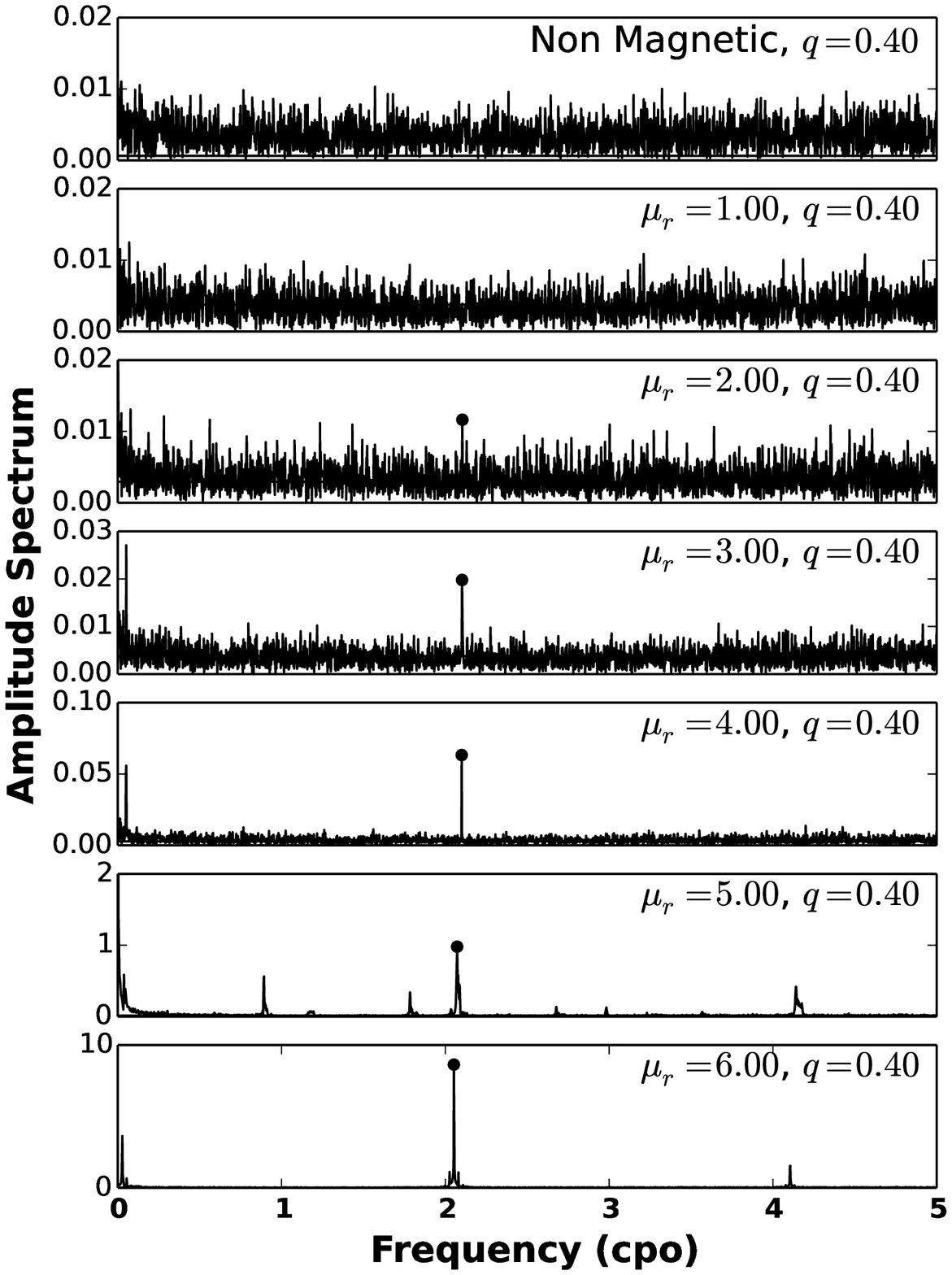}{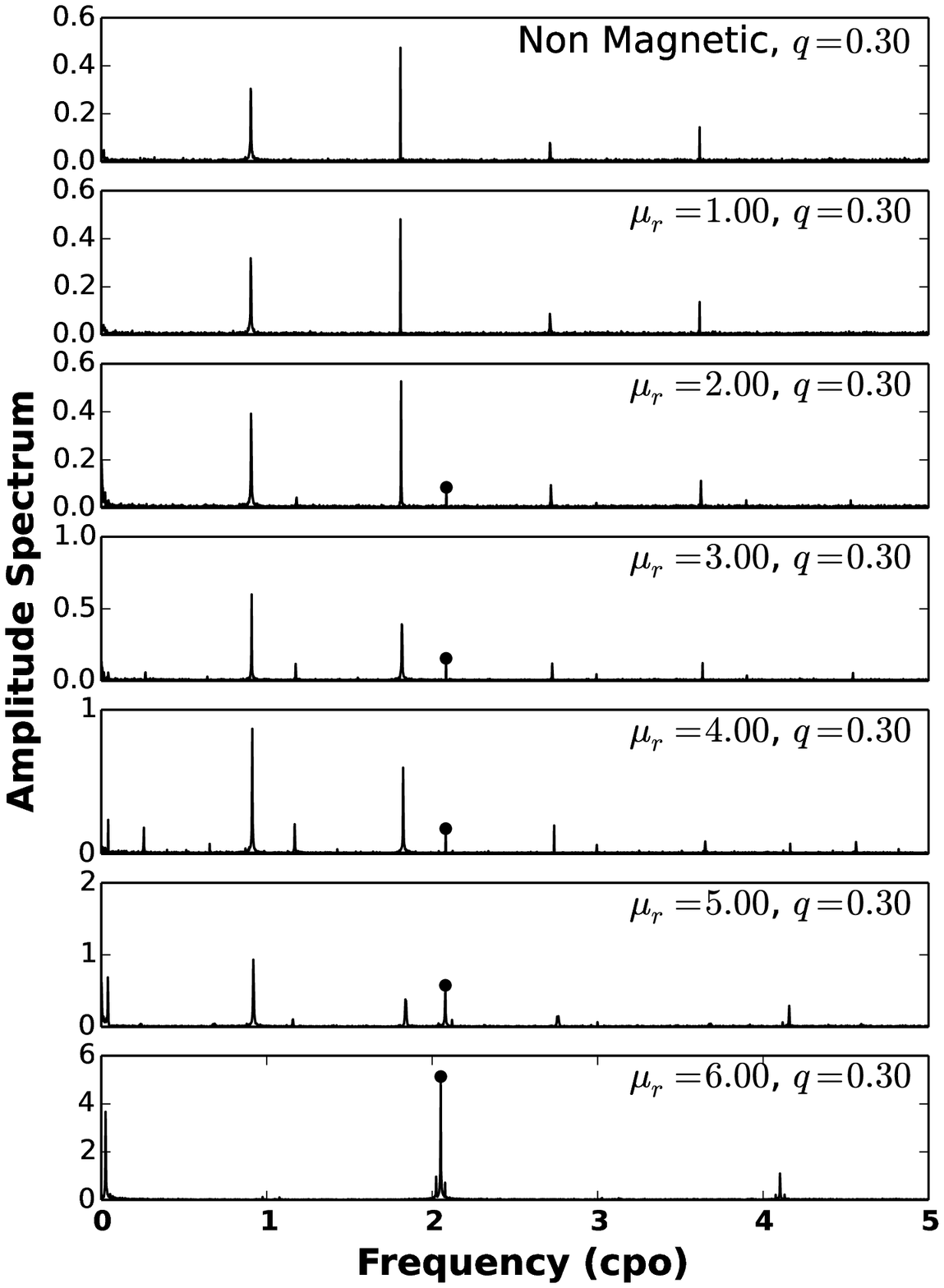}
\caption{Amplitude Spectrum as Scaled Dipole Moment Varies for Two Mass Ratios.  Left Column: $q=0.4$, Right Column $q=0.3$, from top to bottom: Non Magnetic $(\mu_r = 0.0)$, $\mu_r = 1.0$, \ldots, $\mu_r = 5.0$, and (bottom Row) $\mu_r = 6.0$.  Negative superhumps are indicated by filled circles.}
\label{fig:f5}
\end{figure*}

To finish our presentation on the amplitude spectra, we complete the previously presented spectra by extending the range of scaled dipole moments examined.  In Figure~\ref{fig:f5}, the left column corresponds to a mass ratio of $q=0.4$ while the right column represents a mass ratio of $q=0.3$.  As before, we expect the right system to show positive superhumps while the left column should not.  In this Figure, the top row corresponds to a non magnetic system while subsequent rows increase the scaled dipole moments in steps of 1.0.  This Figure shows that for simulation runs that result in the emergence of negative superhumps, the measured superhump frequencies do not change appreciably over a significant range of magnetic field strength.  This conclusion is amplified in Table~\ref{tComp} where we show the variation in the positive and negative superhump frequencies for two mass ratios and the entire suite of scaled dipole moments evaluated in our simulations (the columns labeled $\varepsilon_+$ and $\varepsilon_-$ are defined later).  From Table~\ref{tComp}, we conclude that the negative superhump frequencies change only slightly over the range of magnetic dipoles investigated.  Our results suggest that the magnetic dipole strength provides a mechanism for the emergence of negative superhumps but does not strongly affect the ultimate (frequency) response of the disk.  

\begin{deluxetable*}{cccccccccc} 
\tablecolumns{10} 
\tablewidth{0pc} 
\tablecaption{The Magnetic Dipole Strength has Little Effect on the Negative Superhump Frequency: An Example Showing Two Different Mass Ratios (' \nodata ' represents absent or rejected signals as described in the text).}
\tablehead{
\colhead{} & \multicolumn{4}{c}{$q = 0.3$} & \colhead{} & \multicolumn{4}{c}{$q = 0.4$} \\
\cline{2-5} \cline{7-10} \\
\colhead{$\mu_r $} & \colhead{$\nu_+$} & \colhead{$\varepsilon_+$} & \colhead{$\nu_-$} & \colhead{$-\varepsilon_-$} & & \colhead{$\nu_+$} & \colhead{$\varepsilon_+$} & \colhead{$\nu_-$} & \colhead{$-\varepsilon_-$} \\
\colhead{$ $} & \colhead{(cpo)} & \colhead{(\%)} & \colhead{(cpo)}  & \colhead{(\%)} & & \colhead{(cpo)} & \colhead{(\%)} & \colhead{cpo}  & \colhead{(\%)}}
\startdata 
Non Magnetic &   0.9050 & 10.500 &  \nodata &  \nodata &  & \nodata & \nodata &  \nodata &  \nodata  \\
0.2 &   0.9050 & 10.501 &  \nodata &  \nodata &  & \nodata & \nodata &  \nodata &  \nodata  \\
0.4 &   0.9052 & 10.473 &  \nodata &  \nodata &  & \nodata & \nodata &  \nodata &  \nodata  \\
0.6 &   0.9051 & 10.486 &  \nodata &  \nodata &  & \nodata & \nodata &  \nodata &  \nodata  \\
0.8 &   0.9049 & 10.505 &  \nodata &  \nodata &  & \nodata & \nodata &  \nodata &  \nodata  \\
1.0 &   0.9049 & 10.504 &  \nodata &  \nodata &  & \nodata & \nodata &  \nodata &  \nodata  \\
1.2 &   0.9057 & 10.409 &   2.0891 & -4.266 &  & \nodata & \nodata &  \nodata &  \nodata  \\
1.4 &   0.9054 & 10.446 &   2.0897 & -4.292 &  & \nodata & \nodata &  \nodata &  \nodata  \\
1.6 &   0.9061 & 10.360 &   2.0889 & -4.256 &  & \nodata & \nodata &  \nodata &  \nodata  \\
1.8 &   0.9066 & 10.301 &   2.0883 & -4.230 &  & \nodata & \nodata &  \nodata &  \nodata  \\
2.0 &   0.9069 & 10.267 &   2.0881 & -4.217 &  & \nodata & \nodata &  \nodata &  \nodata  \\
2.2 &   0.9072 & 10.227 &   2.0879 & -4.209 &  & \nodata & \nodata &   2.1062 & -5.041  \\
2.4 &   0.9077 & 10.165 &   2.0875 & -4.189 &  & \nodata & \nodata &   2.1059 & -5.030  \\
2.6 &   0.9084 & 10.086 &   2.0870 & -4.168 &  & \nodata & \nodata &   2.1058 & -5.026  \\
2.8 &   0.9090 & 10.016 &   2.0865 & -4.146 &  & \nodata & \nodata &   2.1053 & -5.000  \\
3.0 &   0.9095 &  9.951 &   2.0861 & -4.126 &  & \nodata & \nodata &   2.1047 & -4.976  \\
3.2 &   0.9102 &  9.863 &   2.0857 & -4.108 &  & \nodata & \nodata &   2.1044 & -4.961  \\
3.4 &   0.9110 &  9.776 &   2.0852 & -4.086 &  & \nodata & \nodata &   2.1039 & -4.939  \\
3.6 &   0.9118 &  9.670 &   2.0847 & -4.064 &  & \nodata & \nodata &   2.1033 & -4.912  \\
3.8 &   0.9127 &  9.570 &   2.0843 & -4.045 &  & \nodata & \nodata &   2.1027 & -4.885  \\
4.0 &   0.9133 &  9.490 &   2.0840 & -4.032 &  & \nodata & \nodata &   2.1020 & -4.853  \\
4.2 &   0.9140 &  9.405 &   2.0838 & -4.020 &  & \nodata & \nodata &   2.1012 & -4.815  \\
4.4 &   0.9148 &  9.316 &   2.0836 & -4.013 &  & \nodata & \nodata &   2.1004 & -4.780  \\
4.6 &   0.9155 &  9.228 &   2.0834 & -4.004 &  & \nodata & \nodata &   2.0926 & -4.427  \\
4.8 &   0.9167 &  9.082 &  \nodata &  \nodata &  & \nodata & \nodata &  \nodata &  \nodata  \\
5.0 &   0.9207 &  8.613 &  \nodata &  \nodata &  & \nodata & \nodata &  \nodata &  \nodata 
\enddata 
\label{tComp}
\end{deluxetable*}

In Figure~\ref{fig:f5}, we see that the scaled dipole moment must be above a specific value for the negative superhumps to appear.  We also see that, as $\mu_r$ increases, there comes a point where sidelobes begin to appear about the frequency of the negative superhump.  In this paper, we accept negative superhumps in the range where a single frequency (without sidelobes) is present and reject the rest.  This selection process leads to results consistent with results obtained previously for manually tilted disks.   The range of $\mu_r$ found to lead to negative superhumps, as determined by our simulations and using our selection process, is presented in Table~\ref{tRange}.

\begin{deluxetable*}{lccccccccccc} 
\tablecolumns{12} 
\tablewidth{0pc} 
\tablecaption{The Range of Scaled Dipole Moments ($\mu_r$) Leading to Negative Superhumps vs. the Mass Ratio ($q$) as obtained by Simulation.}
\tablehead{
\colhead{\ } & \multicolumn{11}{c}{Mass Ratio ($q$)} \\
\cline{2-12} \\
\colhead{$\mu_r$} & \colhead{0.10} & \colhead{0.15} & \colhead{0.20} & \colhead{0.25} & \colhead{0.30} & \colhead{0.35} & \colhead{0.40} & \colhead{0.45} & \colhead{0.50} & \colhead{0.55} & \colhead{0.60}} 
\startdata
Low $\mu_r$ & 3.2 & 1.6 & 1.0 & 1.4 & 1.2 & 1.4 & 2.2 & 2.2 & 1.4 & 1.6 & 1.4 \\ 
High $\mu_r$ & 5.6 & 5.6 & 5.4 & 5.2 & 4.6 & 4.8 & 4.6 & 4.6 & 4.6 & 4.4 & 4.2 
\enddata 
\label{tRange}
\end{deluxetable*}   

\subsubsection{Demonstration of Tilted Disks}

\begin{figure}
\plotone{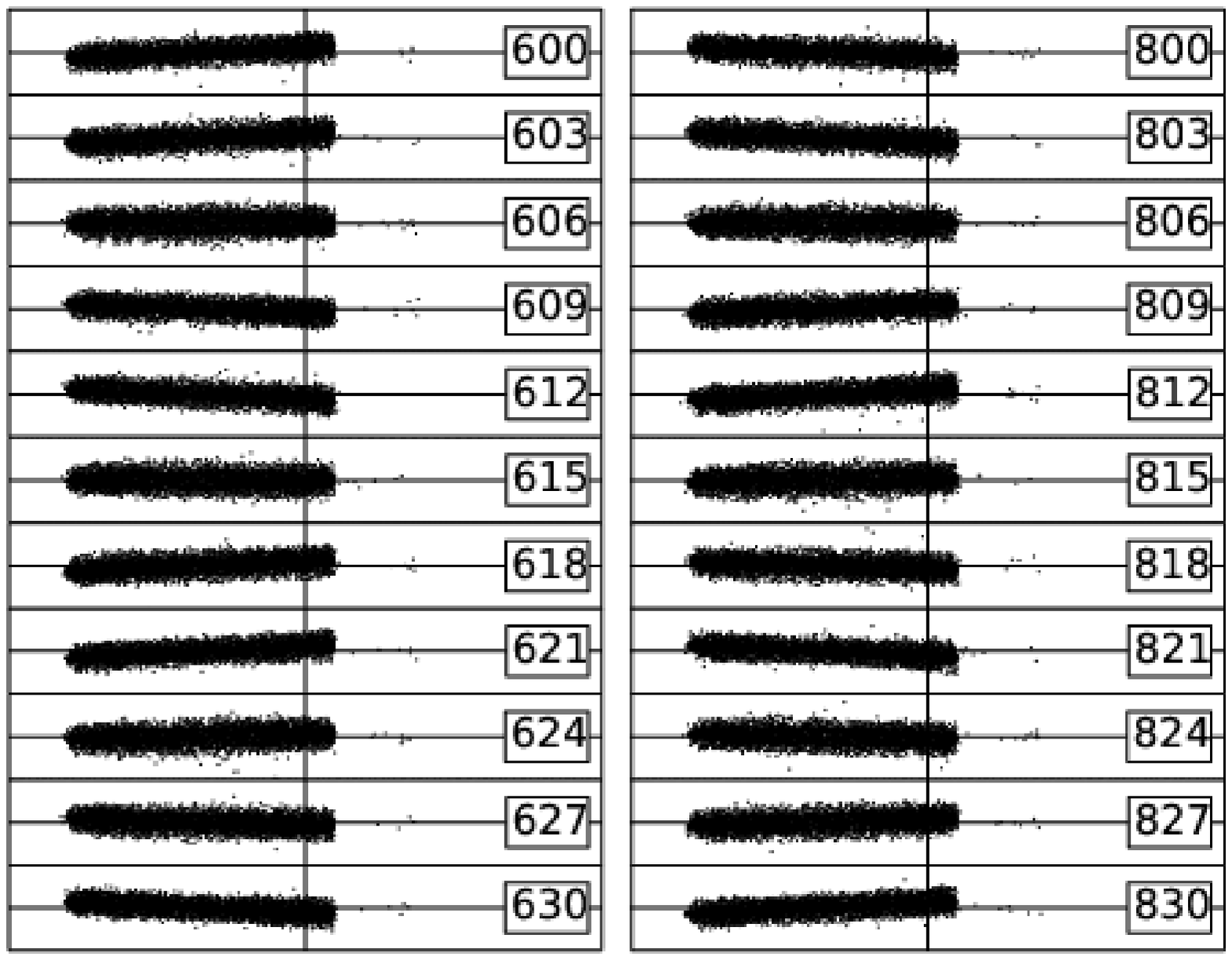}
\caption{Demonstration of disk tilt in a System without Positive Superhumps.  (Left) Orbits 600-630 (from top to bottom in steps of 3 as indicated on the right side of the column) and (Right) orbits 800-830 (also top to bottom) for $\mu_r = 3.0$ and $q = 0.4$.  The vertical scale is exaggerated (horizontal axis ranges over $-0.8a$ to $+0.8a$ while the vertical axis ranges over $-0.05a$ to $+0.05a$).  All plots were pruned by selecting every fifth particle.}
\label{fig:f6}
\end{figure}

\begin{figure}
\plotone{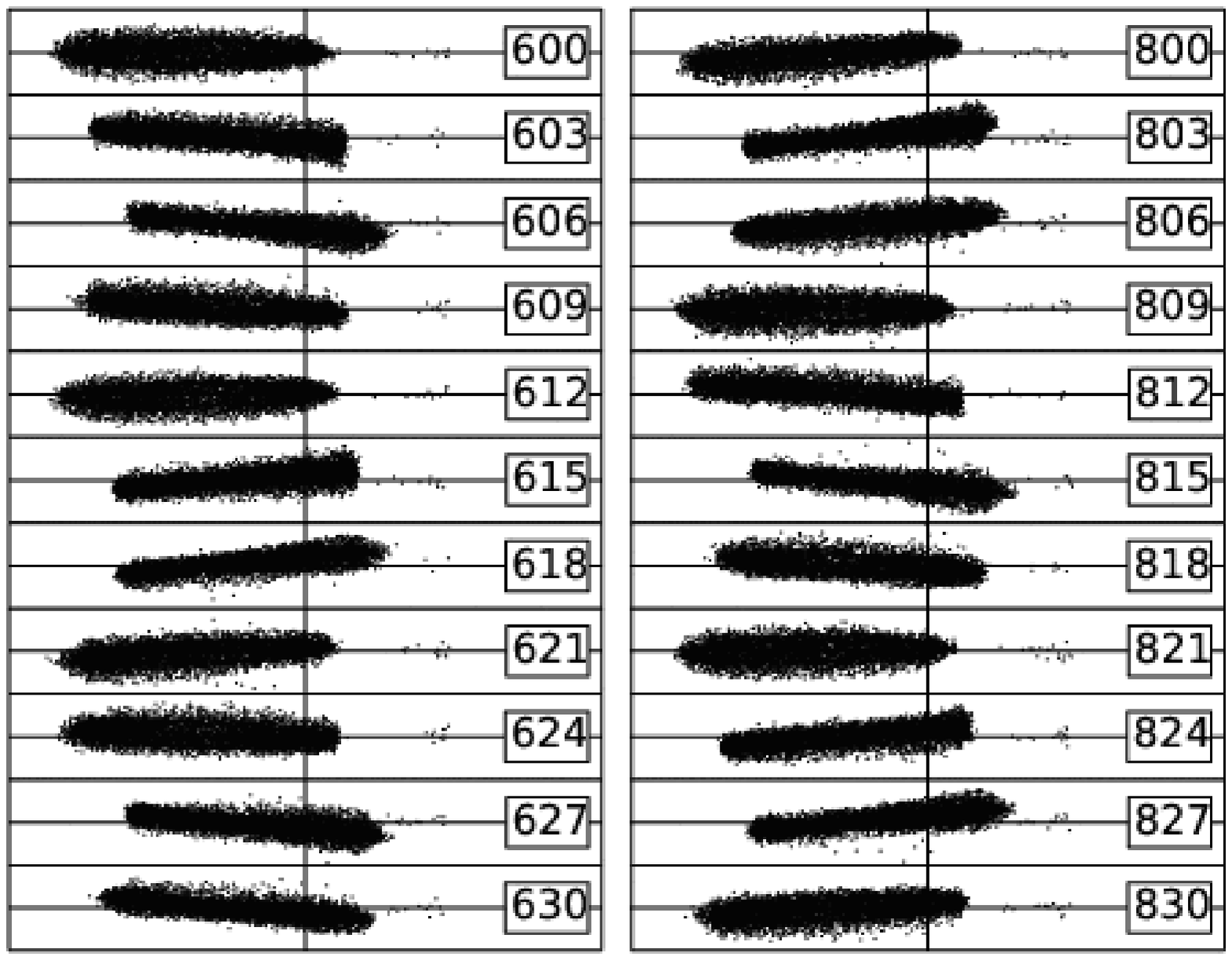}
\caption{Demonstration of disk tilt in a System with Positive Superhumps.  (Left) Orbits 600-630 and (Right) Orbits 800-830 for $\mu_r = 3.0$ and $q = 0.3$.  The vertical scale is exaggerated (relative to the horizontal scale) to emphasize the tilt (horizontal axis ranges over $-0.8a$ to $+0.8a$ while the vertical axis ranges over $-0.05a$ to $+0.05a$).  All plots were pruned by selecting every fifth particle.}
\label{fig:f7}
\end{figure}

We demonstrate disk tilt using the panel plots given in Figure~\ref{fig:f6} for a mass ratio of $q=0.4$ and in Figure~\ref{fig:f7} for a mass ratio of $q=0.3$.  In both plots we set the scaled dipole moment to a value of $\mu_r = 3.0$ and project the particle positions onto the \textit{xz}-plane.  In these plots, the left column of each Figure samples orbits 600 through 630 while the right column samples orbits 800 through 830.  Comparison of Figure~\ref{fig:f7} to Figure~\ref{fig:f6} shows the movement of the disk within the Roche lobe characteristic of systems with positive superhumps.  

We note that the tilt is not constant but varies in time.  Consider the case where $\mu_r = 3.6$ and $q = 0.40$.  Here, after a start up transient, the tilt, defined as the angle between the binary's orbital vector and the disk angular momentum vector, varies with multiple periods within the range of 1.6 to 2.0 degrees.  For the case where $\mu_r = 3.6$ and $q = 0.30$, the tilt varies within the range of 2.9 to 3.7 degrees after the initial transient has disappeared.  Although similar to the previous case, the variation in this instance is more complicated possibly due to the presence of a positive superhump signal.  

The tilt increases with increasing dipole moment.  The tilt varies about a 14 degree center (the variation is about 3 degrees trough to peak) for $\mu_r = 5.0$, $q = 0.3$ and for orbits 750-850 but has not yet achieved steady state.  For orbits 900-1000, the tilt varies about a mean of 15 degrees and appears to be still increasing.  With $\mu_r > 5.0$ and $q = 0.30$, the mean tilt is near 30 degrees varying from 27.5 degrees at the low end to 32 degrees at the high end of the dipole field strength.  In this regime, our simple model likely no longer applies and a more physical magnetohydrodynamic model is likely required.  Therefore, we truncate our Table at $\mu_r = 5.0$.

\subsubsection{Demonstration of Retrograde Precession}

\begin{figure*}
\plotone{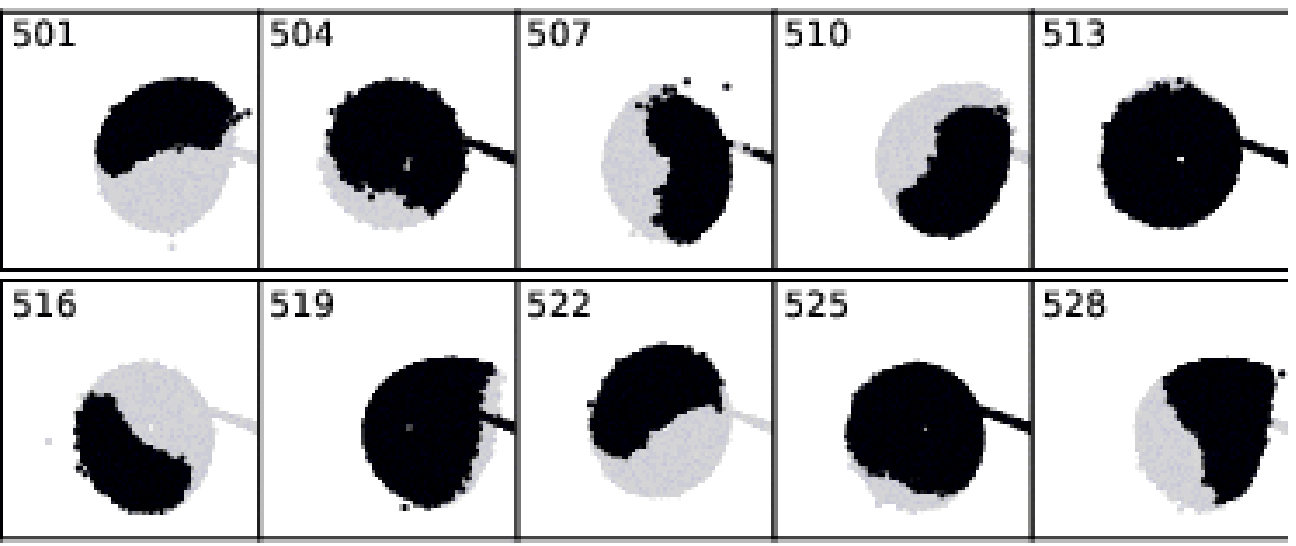}
\caption{Retrograde precession is demonstrated in the case of $q = 0.55$ and $\mu_{r} = 4.6$.  Black is used to color portions of the disk lying above the disk's center of mass in the $z$-direction.  Gray is used to denote particles below the center of mass of the disk.  The orbit is provided in the upper left hand corner (the orbital phase is the same in all cases).}
\label{fig:f8}
\end{figure*}

Snapshops of a system with mass ratio $q=0.55$ and scaled dipole moment $\mu_r = 4.6$ at various orbits (as indicated in the upper left hand corner) are provided in Figure~\ref{fig:f8}.  The dark regions represents portions of the disk's top surface that lie above the disk's center-of-mass in the $z$-direction while gray represents those portion's of the disk's upper surface that lie below the disk's center-of-mass. The small disk flare is revealed by the division between the dark and light regions.

\subsubsection{Period Excess Analysis}
Once the identities of the precession, positive superhump, and negative superhump periods are confirmed, period excesses are calculated.  The period excess is defined as $\varepsilon = (P_{\rm sh} - P_{\rm orb})/P_{\rm orb}$ where $P_{sh}$ is the superhump period and $P_{\rm orb}$ is the orbital period of the binary.  For positive superhumps, the positive period excess is $\varepsilon_+ = (P_+ - P_{\rm orb})/P_{\rm orb}$ where $P_+$ is the positive superhump period.  Similarly, we define the negative superhump deficit as $\varepsilon_- = (P_- - P_{\rm orb})/P_{\rm orb}$ where $P_-$ is the negative superhump period.  Negative superhumps may also be called nodal superhumps or infrahumps while positive superhumps are often called common or apsidal superhumps. The terms apsidal and nodal arise from considerations of physical the origin of the superhumps---precession of the apses of an eccentric oscillating disk or precession of the line-of-nodes of a tilted disk, respectively.  In the case of apsidal superhumps, the precession period is given by $1 / P_{\rm prec}^+ =  1 / P_{\rm orb} - 1 / P_+$.  For nodal superhumps, the precession period is $1 / P_{\rm prec}^- =  1 / P_- - 1 / P_{\rm orb}$.

\begin{figure*}
\begin{center}
\plotone{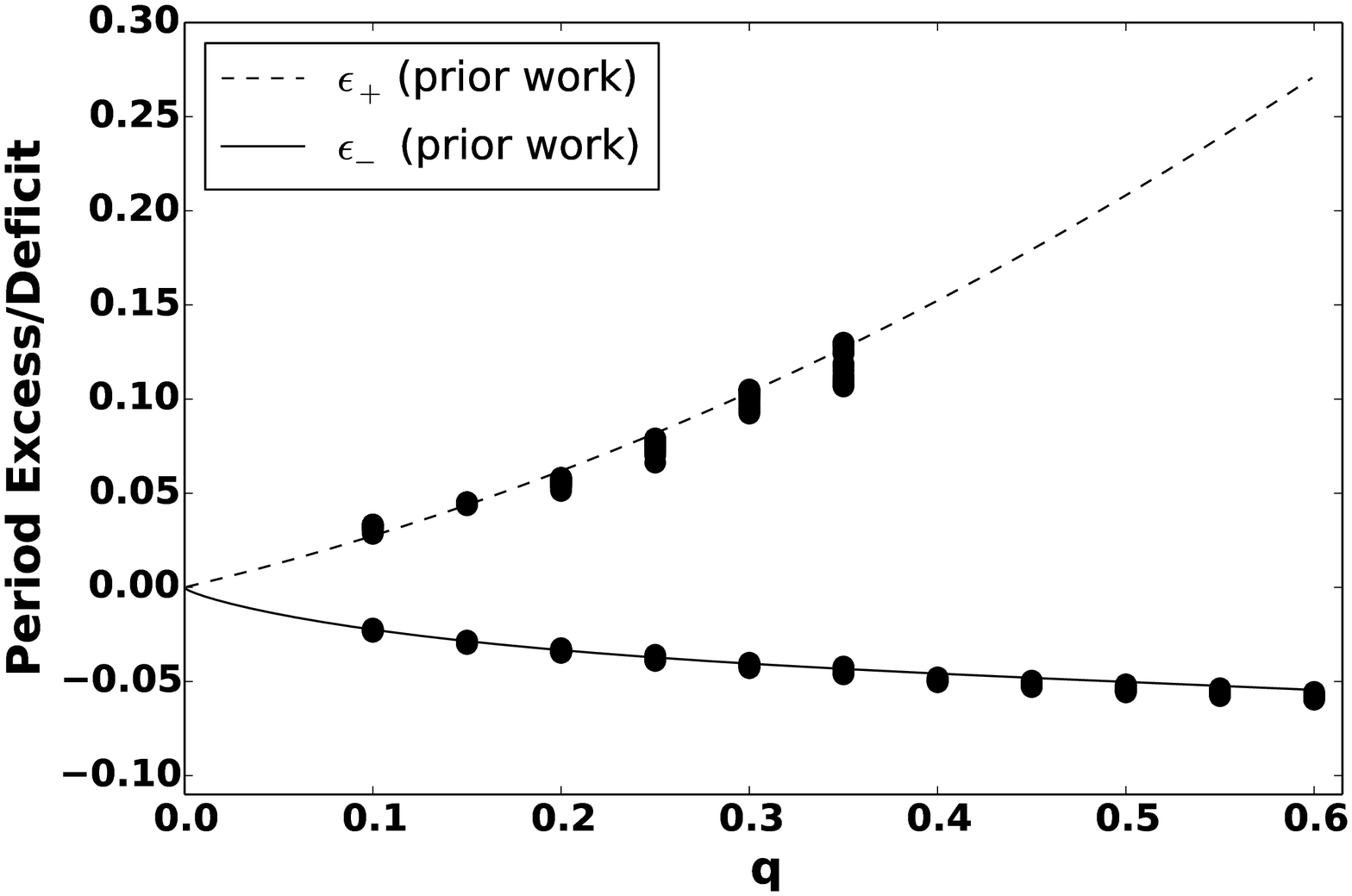}
\caption{Period Excess versus Mass Ratio ($q$) and Scaled Dipole Moment ($\mu_r$) calculated using the last 500 orbits of each simulation. Upper curve is the positive excess while the bottom curve is the negative excess.  Lines are based on fits presented in \citet{2009MNRAS.398.2110W}: $\varepsilon_+ = 0.238 q + 0.357 q^2$ (dashed line) and $\varepsilon_- = -0.02263 q^{1/2} - 0.277 q + 0.471 q^{3/2} - 0.249 q^2$ (solid line).  Individual markers are the results of the simulations described in this paper and include the non magnetic results.  The non-magnetic results fall only on the upper curve (i.e., negative superhumps do not appear in non-magnetic systems.)  All simulations contain 25,000 particles and are continued for 1,000 orbits.}
\label{fig:f9}
\end{center}
\end{figure*}

A plot of period excess versus $q$ and indexed by $ \mu_r$ is provided in Figure \ref{fig:f9}.  In addition, fits to the period excesses obtained from simulations in which the disks were artificially tilted \citep{2009MNRAS.398.2110W} are overlaid on the graphs.  As can be seen the agreement between the two sets of simulations is excellent.  These results also agree with \citep{2012ApJ...745L..25M}, who uses the same basic code. The significance of this plot is that positive superhumps continue to appear even though the code has been modified to allow a magnetic dipole on the primary star.  As such, the upper curve is a validation that we have not seriously affected the apsidal behavior of the disk.  On the other hand, the emergence of negative superhumps and their correspondence to previous results supports the premise of this paper that a magnetic dipole on the primary is, at least, one route to the production of negative superhumps.  The near degeneracy of the data for any given mass ratio highlights once again the possibility that the magnetic field enables the emergence of negative superhumps but does not determine the period.   Our results indicate that the magnetic field can tilt the disk out of the orbital plane, but once tilted, the disk precesses at a rate very near that found in \citep{2009MNRAS.398.2110W}. 

\subsubsection{Some Variations}

The results we have presented to this point demonstrate that the addition of a magnetic field on the primary per \citeauthor{1999ApJ...524.1030L}'s \citeyear{1999ApJ...524.1030L} theory can lead to the appearance of negative superhumps, tilting of disks, and retrograde precession.  Clearly, we have not explored the parameter space to any great depth.  Here, we begin that exploration to see if our results are atypical of what may be expected.

In \citep{2013ARep...57..327B} it is reported that, when the magnetic field on the white dwarf is fixed in the co-rotating coordinate system, negative superhumps are seen to exist but their amplitude decays over time.  To test whether or not our simulations experience a similar fate, we continued the simulations out to 2,000 orbits for two cases ($\mu_r = 3.6$ with $q = 0.3$ or $q = 0.4$.)  For these two cases we determined both the frequency and the amplitude of the negative superhumps for different intervals (namely orbits 501-1000, 1001-1500, and 1501-2000).  No variation in frequency (out to three decimal places) was seen for either mass ratio.  The amplitude of the negative superhump was observed to increase slightly as we moved the analysis window out in time (orbits).  We conclude that the negative superhumps are not decaying.

Next, we examine the possibility that our choice of $\alpha$ in the viscosity prescription can affect the conclusions of this paper.  We ran simulations for $\mu_r = 3.6$ with $q = 0.3$ or $q = 0.4$.  The first viscosity parameter was chosen to be either $\alpha = 1$ or $\alpha = 0.5$.  Virtually no variation was seen in the period of the negative superhump when simulations of equal mass ratios were compared. However, the amplitude of the negative superhump was seen to increase with decreasing $\alpha$.  We conclude that the appearance of negative superhumps is not dependent on the value of the viscosity parameter at least for the range of values examined here.

\subsection{Relation to physical stars}
The scaled dipole moments used in the previous section roughly bound the possible range of values.  Values of $ \mu_r$ smaller than $\simeq 1.0$ do not lead to tilting of the disks nor to nodal superhumps. Use of values above $6.0$, lead to rapidly tilting disks, the appearance of sidelobes, and (in a small number of cases) code termination from numerical issues.  The precise range of values is dependent on $q$ (recall Table~\ref{tRange}) and possibly other parameters.  Accepting for now that this range of $ \mu_r$s is not too far off, we want to determine the equivalent magnetic dipole moments possible in real stars.  To this end, we re-arrange Equation~(\ref{EqMuSq}) and substitute the definitions of $k_r$ and $\mu_r$ to get
\begin{equation}
\mu =  k_{r} \Omega_{orb} a^{2.5625} m_1^{-0.1875} N^{-0.5} \left( \alpha_{SS}^{-0.45} {\dot M_{16}}^{0.425} \mu_r \right),
\nonumber
\label{EqMu}
\end{equation}
\noindent where $k_{r} = \sqrt{k_{r}^2} = 10^{4}\sqrt{k_{\mu}^2} = 1.378 \times 10^{6}$ g$^{0.5}$ cm$^{-0.0625}$.  

In our simulation, $m_1$ and $m_2$ are provided as input and $q$ is calculated.  To complete the description of the system, the user is allowed to choose whether to provide $P_{\rm orb}$, $a$, or use (as we have in all simulations presented in this paper) a mass-radius relationship appropriate to the main sequence companion star \citep{1995ApJ...448..822S}.  In the mass-radius relationship available, the secondary's radius is given by
\begin{equation}
r_s = \left( M_2/M_\sun \right)^{\epsilon} R_{\sun} 
\label{EqMassRadius}
\end{equation}
\noindent where $\epsilon = 13/15 \approx 0.867$ \citet[pg. 194]{2001cvs..book.....H}, $M_2$ is the secondary's mass (g), $M_{\sun}$ is the solar mass (g), and $R_{\sun}$ is the solar radius (cm).  \citet[see Equation 2.5c on page 33]{1995CAS....28.....W} provides an equation, attributed to Eggleton, for the radius of the secondary star $\left(r_s\right)$ relative to $a$.  It is 
\begin{equation}
\frac{r_2}{a} = \frac{0.49q^{2/3}}{0.6q^{2/3}+\ln{\left(1 + q^{1/3}\right)}}.
\end{equation}
Combining this equation with the mass-radius relation given in Equation~(\ref{EqMassRadius}) allows us to determine the stellar separation $a$.  With $m_1$, $q$, and $a$ in hand; we can find, using Kepler's law, the orbital period $\left( P_{\rm orb} \right)$ or, equivalently, the orbital angular frequency $\Omega = 2\pi / P_{\rm orb}$.  Use of preceding parameters and the scaled dipole moment in Equation~(\ref{EqMu}), leads to a physical value for the dipole moment.  

\begin{figure}
\plotone{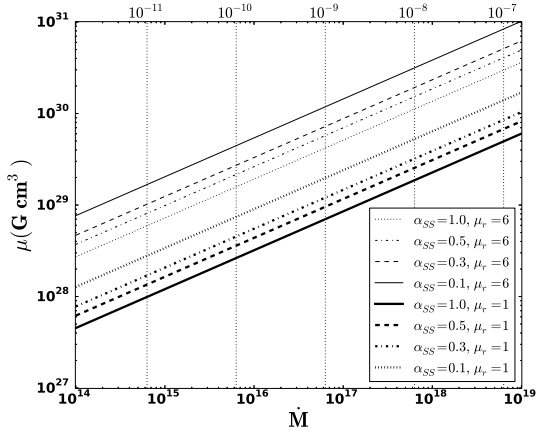}
\caption{Magnetic Moment $\mu$ (G cm$^3$) vs. $\dot M$ for various values of $\alpha_{SS}$ and two values of $\mu_r$: Weak Field ($\mu_r = 1.0$) and a Strong Field ($\mu_r = 6.0$).  The mass ratio is set to $q = 0.4$, and N (i.e., number of particles) is 25,000.  The bottom scale has units of g s$^{-1}$ while the top scale is in units of $M_{\sun}\ yr^{-1}$ and matches the vertical lines.  The text describes the approach used to calculate these results.}
\label{fig:f10}
\end{figure}

Figure~\ref{fig:f10} presents an evaluation of Equation~(\ref{EqMu}) as the accretion rate is varied for a variety of viscosity coefficients and two scaled dipole moments corresponding to a weak field and a strong field.  The vertical lines in this Figure represent, as labeled, accretion rates in solar masses per year.  The five vertical lines given bound the range of expected accretion rates with $10^{-10} M_{\sun}\rm\ yr^{-1}$ representing dwarf novae (DN), $10^{-9} M_{\sun}\rm\ yr^{-1}$  appropriate to Z Cam stars (ZC), and $10^{-8} M_{\sun}\rm\ yr^{-1}$ associated with the nova-likes (NL) as suggested by \citep[See Figure 9.8, pg. 476]{1995CAS....28.....W}.    

Based upon the simulations presented in this paper, we suggest that magnetic dipoles strengths in the region $10^{28}$ to $10^{31}$ G cm$^3$ are required to support the emergence of negative superhumps.  This range lies just below the strengths of magnetic dipole strengths found \citep{2004ApJ...614..349N, 1994PASP..106..209P} in both intermediate polars and polars (namely, $\lesssim 10^{32}$ to $>10^{34}$ G cm$^3$).  Some overlap with the lower portions of the range of magnetic fields found in the intermediate polars is possible given the number of systems listed in the references just given that are reported to have strengths $\lesssim 10^{32}$ G cm$^3$. 

\begin{figure}
\plotone{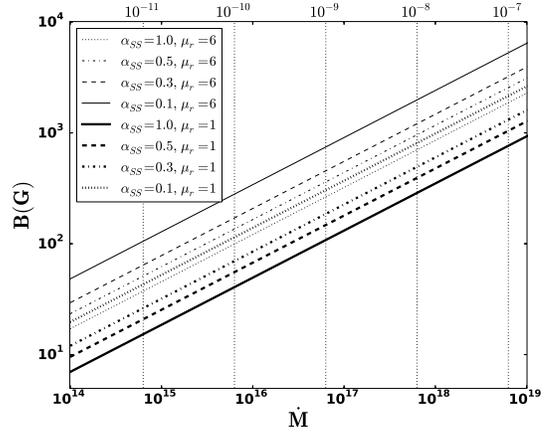}
\caption{Magnetic Field Strength (Gauss) vs. $\dot M$ for various values of $\alpha_{SS}$ and two values of $\mu_r$: Weak Field ($\mu_r = 1.0$) and a Strong Field ($\mu_r = 6.0$). The primary mass is set to $m_1 = 0.6$, the mass ratio is set to $q = 0.4$, and N (i.e., number of particles) is 25,000.  The bottom scale has units of g s$^{-1}$ while the top scale is in units of $M_{\sun}\ yr^{-1}$ and matches the vertical lines.  Magnetic fields in this figure correspond to the magnetic dipole moments presented in Figure~\ref{fig:f10}.  The text describes the approach used to calculate these results.}
\label{fig:f11}
\end{figure}

The magnetic field strength, $B$, at the surface of the primary may be found from $\mu$ using the relation $B = \mu / r_{\rm wd}^3$ where $r_{\rm wd}$ is the radius of the white dwarf.   An approximation for the radius of the white dwarf is given in \citep[page 193]{2001cvs..book.....H} where, 
\begin{equation}
r_{\rm wd} = 7.79 \times 10^{8}\sqrt{M_C^{-2/3} - M_C^{+2/3}} \ \rm {cm,}
\end{equation}
with $M_C$ equal to the ratio of the white dwarf mass to the Chandrasekhar mass $\left(1.44M_{\sun}\right)$.  A graph of magnetic field strengths is given in Figure~\ref{fig:f11} corresponding to the magnetic moments provided in Figure~\ref{fig:f10}. These results are consistent with the magnetic field strengths invoked \citep[i.e., $\mu \gtrsim 5 \times 10^{30}$ G\ cm$^3$]{2002A&A...394..231H} in VY Scl stars as one way (though not their preferred approach) to explain the absence of outbursts during their low state. 

The field strengths determined above are low and the questions arises whether there is any evidence that kilo-gauss fields such as these are found in white dwarf stars.  Isolated magnetic white dwarfs with kilogauss strengths have been reported by \citet{2004A&A...423.1081A} and \citet{2007A&A...462.1097J} using the Very Large Telescope FORS1 instrument.  The possibility that some FORS1 surveys contained spurious detections has been raised by \citet{2012A&A...538A.129B}, \citet{2012A&A...542A..64J}, and \citet{2012A&A...541A.100L}. A subsequent re-evaluation of the data, as provided in \citep{2012A&A...545A..30L} and \citep{2013ASPC..469..411J}, has demonstrated the possibility of white dwarf kilo-gauss magnetic fields.  Unfortunately, the white dwarf does not dominate the light in CVs, and as a result it is unlikely that such observations could be feasible in these systems. 

\subsection{Relation to Previous Work}

Others have also applied Lai's (1999) model, although not in the context of cataclysmic variables.
\citet{sl2002} model the global warping and precession modes in a viscous accretion disk in the context of accretion onto X-ray pulsars.  These are numerical integrations, not hydrodynamic simulations.  These authors find that the inner region of the disk around a strongly magnetized accretion disk can be warped and precess around the stellar spin, and suggest that the effect they find may be responsible for the quasi-periodic oscillation signals observed in some low-mass X-ray binaries.  They do not find that the outer disk also gets tilted as we find, however.
\citet{2004ApJ...604..766P} model the nonlinear evolution of warped, viscous accretion disks resulting from the magnetic torques predicted by Lai (1999).  These authors also find very large-angle tilting of the inner disk, but it is difficult to compare their results with ours as they assume that the outer disk will be unaffected and adopt this as an outer boundary condition.  They note that there models are limited in that they are based on phenomenological descriptions of viscosity, and that their results may depend on the physics at the inner radius of the accretion disk.   
\citet{lai2014} models the dynamics of a circumstellar (proto-planetary) disk from an external binary companion.  In this work, one of his findings is that the tidal torque from the external binary companion makes the circumstellar disk precess as we have found here.  He notes that the disk behaves approximately as a rigid body because adjacent regions are coupled by internal waves, viscosity, or gravity.  We suggest that the viscosity prescription we use (standard for SPH simulations) may result in a higher effective coupling in the disk than that used by the authors above.  The simulations we present here are preliminary -- we are currently developing a new-generation SPH code and plan to explore these effects in more depth.

As noted above, Montgomery (2012a,b) identifies natural tilting of disks as a source of negative superhumps.  She, using the same basic code as we use in this work \citep{1998ApJ...506..360S}, has found that after many hundreds of orbital periods, that the disk tilts on its own without requiring magnetic fields or radiation pressure.  Her model for the mechanism that tilts disks is that the disk behaves like an airfoil, and that it is a lift force originating in the interaction region between the ballistic accretion stream and the disk rim that tilts the disk out of the orbital plane.  In our previous work using the same code (without Lai's magnetic force), we only saw this effect in one simulation, and that was of a very low mass ratio $q\approx0.025$.   Although Montgomery reports a clear signal at $2\nu_-=2.097\rm\ orbit^{-1}$ for her $q=0.35$ simulation, our $q=0.35$ energy generation curves do not show any sign of negative superhumps in the Fourier transform.  However, we used higher viscosity coefficients, $\alpha=1.0, \beta=0.5$, than Montgomery used, $\alpha=0.5, \beta=0.5$.  We plan to run a set of simulations to replicate the Montgomery results and to extend them to higher particle number and other viscosity values, and will report the results of that project in a future publication.  

\subsection{Future Work and Model Enhancements}

These results represent a preliminary investigation that includes only a portion of the available parameter space.  For example, we have not investigated the impact of varying $\zeta$ on the results.  Also, the magnetic field induced precession has not been found to be significant for the range of parameters we have studied.  One possible reason for this is that we apply the force responsible for precession to all particles and not only to those at the surface of the disk.  To be more physically correct, we would need to locate the surface of the disk and apply the appropriate force, perhaps, accounting for skin depth.  Even without this change, it would be interesting to compare the range of magnetic field induced precessions as the parameters of the model are varied to the predictions of tidally induced precession developed by \citet{1995MNRAS.274..987P}.  We can also begin an investigation into the impact of placing the magnetic dipole on the secondary as the code described in this paper has that ability.  Work done early in this research indicated that effects similar to those reported in this paper appear when the secondary hosts the magnetic dipole.  This alternative placement of the dipole would serve to support previous studies (mentioned in the introduction) that address this scenario.  Simple modifications to the code would allow magnetic fields to be placed on both stars.  

There are also a number of enhancements to the code that are possible.  First, we can complete the force model to include the radial and azimuthal force terms ignored in this work.  Another direction we can explore is to evaluate existing physical theory for suggestions on possible functional forms for $\zeta$.  Further, the time varying magnetic field should have a time varying impact on the magnetospheric boundary.  We have not modeled this in our code.  An additional concern is that we must import analytical solutions of the Shakura-Sunyaev thin disk model.  Effort could be spent to investigate the possibility of removing this dependency.  In addition, we can instrument our code to investigate the impact of the spin dependent terms appearing in our force model.  This could be used to support the analysis presented in \citet{2008ApJ...683..949L}. In the area of validation of our results, other codes, such as the one described in \citep{2012MNRAS.421...63R, 2013MNRAS.430..699R}, might be used to validate the results we report here. 

\section{Summary and Discussion}

In this work, we have demonstrated that the addition of a magnetic field (i.e., \citeauthor{1999ApJ...524.1030L}'s \citeyear{1999ApJ...524.1030L} model) to an existing SPH code leads to the emergence of negative superhumps in a natural way with period deficits consistent with results using manually tilted disks.  Our simulations suggest that the magnetic field on the white dwarf, while it enables the disk to tilt yielding retrograde precession of the disk (consistent with theory) and leads to the emergence of negative superhumps, does not determine the superhump period.  We suggest that, in fact, it is the natural precession resulting from the presence of the secondary star that is responsible for the observed period.  

Our results may have added significance given recent work by \citet{2011ApJ...741..105W} and \citet{2013PASJ...65...50O}.  In particular, \citeauthor{2013PASJ...65...50O} report the presence of two types of supercycles in V1504 Cyg labeled, using a notation introduced by \citet{1985AcA....35..357S}, `L' and `S' for long and short.  They report that the type `L' supercycle is associated with the presence of negative superhumps and has fewer normal outburst per supercycle (the quiescent period between outburst is longer) when compared to the type `S' supercycle.  They proposed an explanation for this behavior using a tilted disk and cite additional systems demonstrating this behavior.       

We have also shown a way to relate the simulated value of the dipole moment, namely, the scaled dipole moment, to its equivalent physical value.  The magnetic moments necessary for the creation of negative superhumps were found to vary over the range $10^{28}$-$10^{31}$ G cm$^3$.  Magnetic dipole strengths found \citep{2004ApJ...614..349N, 1994PASP..106..209P} in both intermediate polars and polars vary from less than $\lesssim 10^{32}$ to greater than $10^{34}$ G cm$^3$.  Since the inner portions of the disks in intermediate polars are truncated and disks are non-existent in polars, it seems reasonable to expect that the required strengths necessary for negative superhumps should be lower than or overlap those found in magnetic CVs.  This is the result we have obtained.  Our results are also somewhat weaker than magnetic field strengths suggested \citep[i.e., $\mu \gtrsim 5 \times 10^{30}$ G\ cm$^3$]{2002A&A...394..231H} as one way to suppress outbursts in VY Scl stars during their low state.

Our results also suggest that the occurrence of negative superhumps in a non-IP system can be taken as a strong indication that the primary has a magnetic field that is only somewhat weaker than that required for IP behavior.  These systems should be good candidates for other weak-field magnetic effects such as dwarf nova oscillations, quasi-periodic oscillations, and related phenomena \citep[see][]{warner2004,ww2006,ww2009,hellier2014}.  

For the larger relative dipole moments used in this study, the negative superhump spectral lines show a more complicated structure and the period deficits (and period excess for case where positive superhumps occur) move off the fits based on manually tilted disks.  The magnetic field strengths in these cases begin to reach the levels where intermediate polar or IP behavior is expected.  At this point, the model we present here becomes less descriptive of the physical state of affairs.  As the system becomes more like an IP, the inner boundary of the accretion disk moves farther out from the white dwarf and accretion occurs along the field lines and onto the poles of the central star.  We do not model this behavior.  Since the forces leading to tilt in our model fall off quickly with distance from the white dwarf, it is possible that a natural cessation of negative superhump can occur as the the inner disk's boundary recedes from the accretor.  Our Table~\ref{tRange} may then represent the region where negative superhumps can occur and where (using the upper limit) the transition to IP status can be expected.  More detailed modeling of this transition region, possibly using a more complicated magnetohydrodynamic code, is necessary.       
 
\acknowledgments
This paper is dedicated to the memory of Dr. Jim Simpson who passed away unexpectedly and much too soon on 2013 December 14.  Jim was a friend, colleague, and the author of the SPH code that formed the basis of much of our work over the past many years including the results presented here.  He will be missed. 

This material is based upon work supported by the National Science Foundation under Grant No. AST-11092332 to the Florida Institute of Technology and AST-1305799 to Texas A\&M University-Commerce.

\end{document}